\documentclass[11pt,reqno]{article}

\usepackage{amsmath}
\usepackage{amsfonts}
\usepackage{amssymb}
\usepackage{amsxtra}
\usepackage{graphicx}
\usepackage{caption}
\usepackage[titletoc]{appendix}
\usepackage{ushort}
\usepackage{color}
\usepackage{IEEEtrantools}
\usepackage[hang]{footmisc}
\usepackage{hyperref}
\hypersetup{linktocpage=true}
\usepackage{float}

\makeatletter
\@addtoreset{equation}{section}
\makeatother
\renewcommand{\theequation}{\thesection.\arabic{equation}}
 \def\bd{\begin{document}} \def\ed{\end{document}}
\def\ds{\documentstyle} \let\fr=\frac \let\bl=\bigl \let\br=\bigr
\let\Br=\Bigr \let\Bl=\Bigl
\let\bm=\bibitem
\let\na=\nabla
\let\pa=\partial \let\ov=\overline

\newcommand{\be}{\begin{equation}}
\newcommand{\ee}{\end{equation}}
\newcommand{\bse}{\begin{subequations}}
\newcommand{\ese}{\end{subequations}}
\newcommand{\bea}{\begin{eqnarray}}
\newcommand{\eea}{\end{eqnarray}}
\newcommand{\ba}{\begin{array}}
\newcommand{\ea}{\end{array}}
\newcommand\numberthis{\addtocounter{equation}{1}\tag{\theequation}}

\def\ft#1#2{{\textstyle{{\scriptstyle #1}\over {\scriptstyle #2}}}}
\def\fft#1#2{{#1 \over #2}}
\newcommand{\del}{\partial}
\newcommand{\vp}{\varphi}
\def\sst#1{{\scriptscriptstyle #1}}
\def\oneone{\rlap 1\mkern4mu{\rm l}}
\newcommand{\td}{\tilde}
\newcommand{\wtd}{\widetilde}
\newcommand{\ie}{{\it i.e.\ }}
\newcommand{\cf}{{\it c.f.\ }}
\newcommand{\eg}{{\it e.g.\ }}
\newcommand{\egns}{{\it e.g.}}
\def\dalemb#1#2{{\vbox{\hrule height .#2pt
        \hbox{\vrule width.#2pt height#1pt \kern#1pt
                \vrule width.#2pt}
        \hrule height.#2pt}}}
\def\smsquare{\mathord{\dalemb{6.8}{7}\hbox{\hskip1pt}}}
\newcommand{\ho}[1]{$\, ^{#1}$}
\newcommand{\hoch}[1]{$\, ^{#1}$}
\newcommand{\ra}{\rightarrow}
\newcommand{\lra}{\longrightarrow}
\newcommand{\Lra}{\Leftrightarrow}
\newcommand{\ap}{\alpha^\prime}
\newcommand{\bp}{\tilde \beta^\prime}
\newcommand{\tr}{{\rm tr} }
\newcommand{\Tr}{{\rm Tr} }
\def\0{{\sst{(0)}}}
\def\1{{\sst{(1)}}}
\def\2{{\sst{(2)}}}
\def\3{{\sst{(3)}}}
\def\4{{\sst{(4)}}}
\def\5{{\sst{(5)}}}
\def\6{{\sst{(6)}}}
\def\7{{\sst{(7)}}}
\def\8{{\sst{(8)}}}
\def\9{{\sst{(9)}}}
\def\ten{{\sst{(10)}}}
\def\n{{\sst{(n)}}}
\def\cA{{{\cal A}}}
\def\cF{{{\cal F}}}
\def\tV{\widetilde V}
\def\tW{\widetilde W}
\def\tH{\widetilde H}
\def\tE{\widetilde E}
\def\tF{\widetilde F}
\def\tA{\widetilde A}
\def\im{{{\rm i}}}
\def\tY{{{\wtd Y}}}
\def\ep{{\epsilon}}
\def\vep{{\varepsilon}}
\def\R{\rlap{\rm I}\mkern3mu{\rm R}}
\def\bD{{{\bar D}}}
\def\alp{{{\a'}^3}}
\def\R{\rlap{\rm I}\mkern3mu{\rm R}}
\def\bD{{{\bar D}}}
\def\R{{{\Bbb R}}}
\def\C{{{\Bbb C}}}
\def\H{{{\Bbb H}}}
\def\CP{{{\Bbb C}{\Bbb P}}}
\def\RP{{{\Bbb R}{\Bbb P}}}
\def\Z{{{\Bbb Z}}}
\def\bA{{{\Bbb A}}}
\def\bB{{{\Bbb B}}}
\def\bC{{{\Bbb C}}}
\def\bR{{{\Bbb R}}}
\def\bD{{{\Bbb D}}}
\def\bE{{{\Bbb E}}}
\def\bZ{{{\Bbb Z}}}
\def\Re{{{\frak{Re}}}}
\def\Im{{{\frak{Im}}}}
\def\cosec{{\,\hbox{cosec}\,}}
\def\Gm{{\Gamma_{\!\! -}}}
\def\Gp{{\Gamma_{\!\! +}}}

\def\cosech{{\hbox{cosech}}}
\def\sech{{\hbox{sech}}}

\newcommand{\dIdA}{\ensuremath{\dfrac{\delta I}{\delta A}}}
\newcommand{\dIdB}{\ensuremath{\dfrac{\delta I}{\delta B}}}
\newcommand{\threeR}{\ensuremath{\,^{(3)} R \,}}
\newcommand{\threeS}{\ensuremath{\,^{(3)} S \,}}
\newcommand{\zmax}{\zeta_{\text{max}}}
\newcommand{\half}{\frac{1}{2}}
\newcommand{\dadd}{\Delta A''}
\newcommand{\bTwoTilde}{b_2}%{\tilde{b}_2}
\newcommand{\LM}{L_{\! M}} %symbol for the length scale of the mass M (i.e. 2 G M)

\newcommand{\hlinevspace}{\hline & \\[-1.8ex]}
\newcommand{\hlinevspaceThreeCol}{\hline & & \\[-1.8ex]} 
\newcommand{\tamphys}{\it Center for Theoretical Physics,
Texas A\&M University, College Station, TX 77843, USA}

\newcommand{\mitchell}{\it George P. \& Cynthia W.
Mitchell Institute for Fundamental Physics,\\
Texas A\&M University, College Station, TX 77843-4242, USA}
\newcommand{\umich}{\it Michigan Center for Theoretical Physics,
University of Michigan\\ Ann Arbor, MI 48109, USA}
\newcommand{\upenn}{\it Department of Physics and Astronomy,
University of Pennsylvania, Philadelphia,  PA 19104, USA}
\newcommand{\SISSA}{\it  SISSA-ISAS and INFN, Sezione di Trieste\\
Via Beirut 2-4, I-34013, Trieste, Italy}

\newcommand{\newton}{\it Isaac Newton Institute for Mathematical Sciences,\\
20 Clarkson Road,  University of Cambridge,
Cambridge CB3 0EH, UK}

\newcommand{\ihp}{\it Institut Henri Poincar\'e\\
  11 rue Pierre et Marie Curie, F 75231 Paris Cedex 05}

\newcommand{\damtp}{\it DAMTP, Centre for Mathematical Sciences,
 Cambridge University,\\  Wilberforce Road, Cambridge CB3 OWA, UK}
\newcommand{\itp}{\it Institute for Theoretical Physics, University of
California\\ Santa Barbara, CA 93106, USA}

\newcommand{\imperial}{\it The Blackett Laboratory, Imperial College London,\\
Prince Consort Road, London SW7 2AZ, UK }

\newcommand{\beijingnormal}{\it Center for Advanced Quantum Studies, 
Department of Physics, Beijing Normal University,
Beijing 100875, China}

\newcommand{\auth}{
H. L\"u\,\footnote{mrhonglu@gmail.com}\hoch{\dagger},
A. Perkins\,\footnote{a.perkins12@imperial.ac.uk}\hoch{\star},
C.N. Pope\,\footnote{pope@physics.tamu.edu}\hoch{\ddagger} and K.S. Stelle\,\footnote{k.stelle@imperial.ac.uk}\hoch{\star}}

\begin{document}
\pagenumbering{roman}
\setcounter{page}{0}
\thispagestyle{empty}

\begin{flushright}
Imperial/TP/15/KSS/02 \ \ MI-TH-1528
\end{flushright} 

\begin{center}  

\scalebox{.98}[1]{\Large {\bf Spherically Symmetric Solutions in Higher-Derivative Gravity}}   

\vskip 7pt

\auth

\vskip 5pt

{\hoch{\dagger}\beijingnormal} 

\vskip 5pt

{\hoch{\star}\imperial} 

\vskip 5pt

{\hoch{\ddagger}\mitchell}

\vskip 5pt

{\hoch{\ddagger}\damtp}

\vskip 8pt

\underline{ABSTRACT}
\end{center} 
 
Extensions of Einstein gravity with quadratic curvature terms in the action arise in most effective theories of quantised gravity, including string theory. This article explores the set of static, spherically symmetric and asymptotically flat solutions of this class of theories. An important element in the analysis is the careful treatment of a Lichnerowicz-type `no-hair' theorem. From a Frobenius analysis of the asymptotic small-radius behaviour, the solution space is found to split into three asymptotic families, one of which contains the classic Schwarzschild solution. These three families are carefully analysed to determine the corresponding numbers of free parameters in each. One solution family is capable of arising from coupling to a distributional shell of matter near the origin; this family can then match on to an asymptotically flat solution at spatial infinity without encountering a horizon. Another family, with horizons, contains the Schwarzschild solution but includes also non-Schwarzschild black holes. The third family of solutions obtained from the Frobenius analysis is nonsingular and corresponds to `vacuum' solutions. In addition to the three families identified from near-origin behaviour, there are solutions that may be identified as `wormholes', which can match symmetrically on to another sheet of spacetime at finite radius.

\vfill\leftline{}\vfill
\pagebreak

\tableofcontents
\addtocontents{toc}{\protect\setcounter{tocdepth}{2}}
\newpage 
\pagenumbering{arabic}
\setcounter{page}{1}
\setcounter{footnote}{0}

\section{Introduction: Second plus fourth-order gravity }\label{sec:intro}

The inclusion of quadratic curvature terms into the gravitational action is principally motivated by the form of one-loop quantum corrections \cite{'tHooft:1974bx}. In 4D spacetime there are effectively only two independent quadratic-curvature integrated invariants, owing to the existence of the Gauss-Bonnet topological invariant.
Starting from the correspondingly general second-plus-fourth-order action\,\footnote{Using the 4D Gauss-Bonnet theorem \eqref{GBinv}, the action may also be written as
$$\int d^4\sqrt{-g}(\gamma R - 2\alpha R_{\mu\nu}R^{\mu\nu} + (\beta+{2\alpha\over3})R^2)\,.$$}
%%%%%
\be
\label{HDGaction}
I = \int d^4x\sqrt{-g}\left(\gamma R -
\alpha C_{\mu\nu\rho\sigma}C^{\mu\nu\rho\sigma} + \beta R^{2}\right)\,,
\ee
%%%%%
(in which $C_{\mu\nu\rho\sigma}$ is the Weyl tensor, \ie the 
traceless part of  the curvature tensor $R_{\mu\nu\rho\sigma}$),
one obtains\,\footnote{Strictly speaking, for renormalisability one should also include a cosmological constant in \eqref{HDGaction}. Note that the parametrisation of the higher-derivative terms in the action \eqref{HDGaction} differs from that used in Refs \cite{Stelle:1976gc,Stelle:1977ry}. Specifically, $\alpha_{\mbox{here}}=\frac12\alpha_{\mbox{Ref.\,\cite{Stelle:1976gc}}}$ and $\beta_{\mbox{here}}=\beta_{\mbox{Ref.\,\cite{Stelle:1976gc}}}-\frac13\alpha_{\mbox{Ref.\,\cite{Stelle:1976gc}}}$.} a renormalisable system \cite{Stelle:1976gc}. The spectrum of this theory contains \cite{Stelle:1977ry} a massless graviton, a massive spin-two ghost excitation with $(m_2)^2={\gamma\over2\alpha}$, and a massive non-ghost spin-zero excitation with $(m_0)^2={\gamma\over6\beta}$. The canonical value of $\gamma$ is ${1\over16\pi G}={2\over\kappa^2}$, where $G$ is the 4D Newton constant.

The renormalisable quantum system \eqref{HDGaction} is also asymptotically 
free \cite{Fradkin:1981hx,Fradkin:1981iu} in the sense that if one writes 
the coefficients of the quadratic-curvature terms in Yang-Mills style 
as $1/g_2^2$ and $1/g_0^2$, then both couplings $g_2$ and $g_0$ tend to 
zero at large energies. This raises the question as to 
whether the high-energy regime of the model \eqref{HDGaction} might avoid the problems associated with the spin-two ghost in the spectrum by effectively decoupling that excitation at high energies. Such issues have recently been discussed in the context of the asymptotic safety program for quantum gravity \cite{Niedermaier:2006wt}, but to date there does not appear to be a consensus on this point. A key problem in this approach is to obtain robust results that are not renormalization-scheme dependent. A related question, already at the classical level, is whether the interaction structure of the theory might even be such as to avoid the destabilisation of the vacuum by  ghost-driven instabilities \cite{Smilga:2004cy}.

Gravitational theories including quadratic curvature terms arise generically in all approaches to quantum gravity. In particular, the Gauss-Bonnet combination
\be
I_{\rm GB}=\int\sqrt{-g}(R_{\mu\nu\rho\sigma}R^{\mu\nu\rho\sigma}-4R_{\mu\nu}R^{\mu\nu}+R^2)\label{GBinv}
\ee
is a topological invariant in four spacetime dimensions, but not in higher dimensions, where it falls into the class of Lovelock terms \cite{Lovelock:1971yv}. It occurs in the quantum effective action of heterotic string theory in 10 spacetime dimensions \cite{Zwiebach:1985uq}. Various styles of dimensional compactification of \eqref{GBinv} can then yield the quadratic terms of \eqref{HDGaction} in a variety of combinations.\footnote{In the process of dimensional reduction, various massless scalar fields are generated which combine with the $D=10$ dilaton. When the dimensionally reduced $D=4$ theory is written in Einstein frame, scalar field prefactors appear in front of the curvature-squared terms in the effective action.  The study of string-generated higher-derivative gravity models accordingly requires consideration of such scalars together with the quadratic curvature terms. In this paper, however, we shall restrict attention to purely geometric terms in the action.} In dimensions $D>4$, the Lovelock-Gauss-Bonnet combination \eqref{GBinv} also allows for cosmological solutions \cite{Boulware:1985wk}. However, in order to keep our considerations clearly focused, we shall restrict our attention in this paper purely to four-dimensional spacetime gravity derived from \eqref{HDGaction} without a cosmological constant.

In this paper, we will not be concerned with difficult questions of the full physical acceptability of the theory \eqref{HDGaction} at the quantum level. Instead, we shall adopt a working assumption that, in whatever emerges as an acceptable quantum theory of gravity, the system \eqref{HDGaction} may be a dominant part of the effective action at least for some ultraviolet scale of energies. This might have, for example, cosmological implications, which could in turn indicate a scale for the quadratic-curvature term coefficients. It might also be the case that the effects of the quadratic-curvature terms in \eqref{HDGaction} are also characteristic of those of yet higher-order terms. Whatever the fate of the negative-energy massive spin-two excitation, we shall adopt the point of view that its effect on static classical solutions should nonetheless be considered. Accordingly, we shall adopt the action \eqref{HDGaction} as is, and shall consider the implications of its field equations for spherically-symmetric static solutions. We shall thus treat the fourth-order  terms on an equal footing with the second-order terms, and not just as perturbations to the Einstein theory.

Some aspects of the classical solutions to the second-plus-fourth-order gravity theory are well-known. In Ref.\ \cite{Stelle:1977ry}, an analysis was given of spherically-symmetric solutions in the linearised limit of the theory \eqref{HDGaction} when coupled to point and extended sources. As one can expect from a theory whose dynamical spectrum involves massive spin-two and spin-zero modes as well as the massless spin-two Einstein mode, the static solutions to the linearised theory involve both a ${1\over r}$ potential arising from massless spin-two virtual particle exchange and ${e^{-mr}\over r}$ Yukawa potentials arising 
from $m=m_2$ massive spin-two and from $m=m_0$ massive spin-zero 
virtual exchanges. Moreover, by writing the spherically-symmetric and static spacetime metric in Schwarzschild form
\be\label{eq:metric}
ds^{2}=-B(r)\, dt^{2}+A(r)\, dr^{2}+r^{2}d\theta^{2}+r^{2}\sin^{2}\theta\: d\phi^{2}\,,
\ee
assuming a Laurent expansion of $A(r)$ and $B(r)$, 
and carrying out a Frobenius-method analysis of the indicial equations for the leading asymptotic behaviour as $r\rightarrow0$ in the radial coordinate $r$, it was found in Ref.\,\cite{Stelle:1977ry} that the leading asymptotic behaviours\,\footnote{In context, there should be no confusion between the indicial exponents $t$ and $s$ here and the coordinate $t$ and interval $s$ in \eqref{eq:metric}.} $A(r)\sim r^s$ and $B(r)\sim r^t$ can arise in three distinct solution families: $(s,t)=(2,2),\,(1,-1)\ \mbox{or}\ (0,0)$. At the time,  an initial analysis of the number of free parameters characterising these indicial families was made, but in a pre-computer-algebra era, the full picture of such parametric dependences was not easily to be found.

In this paper, we return to a detailed study of the spherically-symmetric solutions to the field equations following from the action \eqref{HDGaction}. Although the classic Schwarzschild solution of Einstein's theory (which belongs to the $(1,-1)$ family) clearly remains a solution to the higher-derivative theory derived from the action \eqref{HDGaction}, we shall find that this is {\em not} a solution that arises from normal minimal coupling to ordinary ghost-free matter. Instead, we find that solutions that can arise from such ghost-free matter coupling belong to the $(2,2)$ indicial family of solutions. Subject to the additional assumption of asymptotic flatness as $r\rightarrow\infty$ at spatial infinity, we find that such solutions do not have a horizon, but have a naked singularity as $r\rightarrow0$. 
This agrees fully, moreover, with numerical calculations of such solutions made in the case $m_2=m_0$ in Ref.\ \cite{Holdom:2002xy}.

If one overlooks the issue of source coupling, which in any case has been a delicate subject in general relativity for decades \cite{Arnowitt:1960zzb,Geroch:1987qn}, then the `black hole' solution family including a horizon can be investigated in its own right. Assuming in addition asymptotic flatness at spatial infinity, the analysis is made much simpler by a Lichnerowicz-style 'no-hair' theorem \cite{Nelson:2010ig} for the trace of the higher-derivative field equations, 
which implies that the existence of a horizon together with the assumption 
of asymptotic flatness leads to the requirement that the Ricci scalar must vanish: $R=0$. Analysis of the remaining traceless components of the field equations is more subtle. In an earlier paper \cite{Lu:2015cqa}, we reported our disagreement with the traceless-equation analysis of Ref.\ \cite{Nelson:2010ig}, which would have considerably simplified the study of the black-hole family. In the absence of a traceless-equation Lichnerowicz theorem, an alternative option
is to make a perturbative analysis of the black-hole family of solutions starting from the classic Schwarzschild solution. We obtain in this way a result that the Schwarzschild solution is at least generally {\em isolated}, in the sense that, for spherically-symmetric static solutions possessing a horizon, solutions perturbatively different from Schwarzschild necessarily must violate the condition of asymptotic flatness at spatial infinity. 

The general perturbative isolation of the Schwarzschild solution within the indicial $(1,-1)$ solution family does not exclude the possibility of other asymptotically-flat and spherically-symmetric solutions with horizons that differ from Schwarzschild by a non-infinitesimal 
amount in the $(1,-1)$ family parameter governing the `non-Schwarzschild' structure of the solutions. Indeed, in Ref.\ \cite{Lu:2015cqa} we demonstrated that this possibility is indeed realised: there exists a range of values for the black-hole horizon radius $r_0$, bounded below by a certain multiple of the $1/m_2=\sqrt{2\alpha/\gamma}$ length scale, for which one obtains a single static black-hole solution in addition to the Schwarzschild solution. The corresponding existence of a minimum value for $r_0$ in comparison to the $\sqrt{\alpha/\gamma}$ scale size in the perturbative no-hair theorem dovetails with the numerically found existence of a branch point for black-hole solution phases. As one approaches this branch point, clearly the perturbative isolation of the Schwarzschild solution must break down.

We begin in Section \ref{sec:equationsOfMotion} with a review of the 
structure of the gravitational field equations following from the 
action \eqref{HDGaction} when restricted to the case of spherically-symmetric and static solutions, initially without considering contributions from sources. In particular, we discuss the reduction of the differential order of these `almost vacuum' equations to get a better fix on the maximum number of integration-constant parameters determining a particular solution family. We shall find that such `almost vacuum' equations reduce to a pair of third-order ordinary but coupled and quite nonlinear differential equations for $A(r)$ and $B(r)$. The full details of these equations are given in Appendix \ref{app:fullequations}. Next, in Section \ref{sec:solutionsNearTheOrigin}, we complete the analysis of the parametric dependence of the various indicial solution families  begun in Ref.\,\cite{Stelle:1977ry}; the advent of Mathematica now makes this much more tractable.

Given that one is principally interested in solutions that are asymptotically flat as $r\rightarrow\infty$, in which limit a linearised analysis of the solution families becomes appropriate, in Section \ref{sec:weakFieldExpansion} we next consider the spherically-symmetric static solutions to the field equations when linearised in $A(r)$ and $B(r)$. In part, this reviews the linearised solutions found already in Ref.\,\cite{Stelle:1977ry}, but with a key addition: we now consider in some detail the matching between an interior vacuum and the exterior solution when matching across a shell delta-function source. 

Coupling to shell delta-function sources in the full nonlinear theory is next taken up in Section \ref{sec:nonlinshellsources}. This discussion begins with a review, in our Schwarzschild-form variables, of the classic analyses of delta-function sources of Refs \cite{Arnowitt:1960zzb,Geroch:1987qn}. In the full nonlinear higher-derivative  theory, exact solutions are not known and so one must use perturbative expansions within the various Frobenius indicial families in order to analyse coupling to a delta-function shell source. Identifying the `vacuum' with the non-singular $(0,0)$ indicial family and requiring this to be the solution type occurring inside a shell source, we find that only the $(2,2)$ indicial family has the correct number of free parameters required to match the various continuity and jump conditions needed across the delta-function source.

Linking what happens near the origin to the behaviour of solutions near spatial infinity becomes the next issue to be considered.  In Section \ref{sec:tracenohair}, we generalise the result of Ref.\ \cite{Nelson:2010ig} to show that the Ricci scalar in a portion of spacetime with Minkowski signature must vanish for any asymptotically-flat solution in the $(0,0)$ or $(1,-1)$ indicial families. For these indicial families, this result obtains regardless of whether one considers a solution with a horizon at some intermediate radius $r_0$, as in Ref.\ \cite{Nelson:2010ig}, or considers a solution without a horizon. Requiring $R=0$ correspondingly reduces the number of free parameters by one in each of these $(0,0)$ or $(1,-1)$ indicial cases.

For the traceless part of the higher-derivative field equation, the situation is complicated by errors made in the analysis of Ref.\,\cite{Nelson:2010ig}, as reported previously in Ref.\ \cite{Lu:2015cqa}. Details of the corrected calculation are given here in Appendix \ref{app:Nelson}. One consequently does not have a straightforward way to prove a complete no-hair theorem setting the full Ricci tensor to zero in the $(0,0)$ or $(1,-1)$ cases. However, for asymptotically-flat solutions with a horizon, one can still use linearised perturbation theory starting from the Schwarzschild solution. First, in Section \ref{sec:horizonexp}, we use a Frobenius analysis about the horizon to show that such solutions, subject also to the requirement of a vanishing Ricci scalar as found in Section \ref{sec:tracenohair}, have just three free parameters. This parameter count identifies the corresponding solution family with the indicial $(1,-1)$ family near the origin, subject also to the requirement of asymptotic flatness as $r\to \infty$ and hence requiring also a vanishing Ricci scalar. The classic Schwarzschild solution is of course itself a member of this family, with just two free parameters (corresponding to the horizon radius and to a trivial time-rescaling parameter). Accordingly, the higher-derivative theory admits just one `non-Schwarzschild' parameter controlling deviations from the Schwarzschild solution.

Deriving a perturbative no-hair theorem for solutions expanded to linear order in the non-Schwarzschild parameter is then carried out in 
Section \ref{sec:Schlinnohair}. For a given horizon radius $r_0$, and for asymptotically-flat solutions treated to linear order in the non-Schwarzschild parameter, one finds that there is a range of small values of $\alpha/(\gamma r_0^2)$ for which the only asymptotically-flat solution with a horizon is the Schwarzschild solution itself. The precise range of such values depends on optimisation details of the linearised no-hair theorem, but the range boundary turns out to be quite near the phase bifurcation point for non-Schwarzschild black hole solutions found numerically in Ref.\ \cite{Lu:2015cqa}. It may turn out that the range boundary for the linearised no-hair theorem and the black-hole phase bifurcation point actually coincide.

In Section \ref{sec:numanalysis}, we consider the more difficult question of what happens more generally in between the origin and spatial infinity. Owing to the complexity of the field equations, this can only be approached by numerical methods. 
One family of solutions at the origin that can mesh with the structures found at spatial infinity is the $(2,2)$ family. 
This agrees with numerical results found  in Ref.\,\cite{Holdom:2002xy} for the specific theory with $m_2=m_0$ (\ie the theory with $\alpha=3\beta$). Generically, such $(2,2)$ solutions have six free parameters at the origin, and six parameters at infinity, of which two combinations in each set must be adjusted in order to kill rising exponential behaviour from the spin-two and spin-zero sectors of the theory, thus leaving a four free-parameter set at the origin corresponding to the four parameters occurring at asymptotically-flat spatial infinity. We give another illustration of such a solution for the $\gamma R-\alpha C^2$ theory (\ie with $\beta=0$), in which the equations simplify owing to the absence of the spin-zero mode, then displaying a restricted match between a three-parameter set at the origin and a three-parameter set at asymptotically-flat spatial infinity. These $(2,2)$ solutions cannot have horizons, since we have established in Sections \ref{sec:tracenohair} and \ref{sec:horizonexp} that asymptotically-flat solutions with horizons must belong to the $(1,-1)$ family. Instead, asymptotically-flat solutions displaying Yukawa massive corrections at spatial infinity track closely to the Schwarzschild solution far out from the radius where the Schwarzschild horizon would have been, but they then begin to differ strongly from Schwarzschild as one comes in toward smaller radii, failing to have a horizon but matching instead onto the $(2,2)$ indicial family of solutions near the origin, and displaying a naked singularity.

Section \ref{sec:numanalysis} also considers the structure of the $(1,-1)$ solution family. The conclusion one draws from the linearised no-hair theorem of Section \ref{sec:Schlinnohair} is that the Schwarzschild solution is generally {\em isolated} within the family of $(1,-1)$ asymptotically-free solutions with a horizon, except for values of $\zeta=\alpha/(\gamma r_0^2)$ located above a certain value $\zeta_{\text{max}}$, which presumably may be identified with the black-hole phase bifurcation point. Below this bifurcation point, perturbation in the single non-Schwarzschild parameter away from the Schwarzschild solution within the $(1,-1)$ family initially can only lead to non-asymptotically-flat solutions. The linearised no-hair theorem is thus in full agreement with the conclusions found numerically in Ref.\ \cite{Lu:2015cqa}. The $(1,-1)$ family naturally contains the Schwarzschild solution itself, with non-asymptotically-flat solutions generally occurring nearby as one adjusts the non-Schwarzschild parameter. As found in Ref.\ \cite{Lu:2015cqa}, however, there do exist additional asymptotically-flat $(1,-1)$ solutions with a horizon that in general must be distinctly separated from the Schwarzschild solution in the value of the non-Schwarzschild parameter. Such non-Schwarzschild solutions occur in the $\zeta<\zeta_{\text{max}}$ range for which the perturbative no-hair theorem is applicable. This other branch of asymptotically-flat black-hole solutions accordingly exists for horizon radii $r_0$ greater than a certain value $r_0^{\rm min}$. At spatial infinity, such non-Schwarzschild black holes have a $\ft1{r} e^{-m_2 r}$ Yukawa correction to the $g_{00}$ metric component in addition to the $2M/r$ Newtonian term, where $M>0$ is the ADM mass. 

Numerical study using the shooting method for the horizonless $(2,2)$ solutions and for the non-Schwarzschild black-hole solutions reveals another feature of the overall spherically-symmetric and asymptotically-flat solution space. Such solutions, with well-understood behaviours in each (small $r$ and large $r$) asymptotic region of the radial coordinate $r$, appear to lie on separatrices between numerically found solutions with differing kinds of divergent behaviour. The implications of this separatrix structure for the overall solution space remain to be more fully understood.

Another type of asymptotically-flat solution that emerges from numerical study may be described as a `wormhole'. In such a solution, which we also discuss in Section \ref{sec:numanalysis} for the $\beta=0$ theory, the inverse of the $A=g_{rr}$ component of the metric goes to zero but the $-B=g_{tt}$ component does not. General $\Z_2$ symmetric solutions of this type are highly constrained, with only two free parameters: the trivial time-rescaling parameter and the radius $r_0$ at which $1/A$ vanishes. Numerical results show that such solutions can achieve asymptotic flatness at spatial infinity only for a particular value of $r_0$, which is presumably related to the $\sqrt{2\alpha/\gamma}$ length scale. The $\Z_2$ symmetric wormhole solution is also found to lie on a separatrix lying between less regular solutions.

In the Conclusion (Section \ref{sec:conclusion}) we give a brief discussion of some possible physical implications of our results. Clearly, the physical relevance of the present analysis depends upon fully accepting the implications of the higher-derivative terms in the field equations for the theory's solutions, instead of simply considering their effects as perturbations on the second-order Einstein theory. It is equally important that there be at least some range of energy/length scales at which the fourth-order terms dominate, without their being swamped by the effects of yet higher-order terms. Given such assumptions, we comment on stability questions for the various spherically-symmetric solutions, and on the possibility of phases in which the classic Schwarzschild solution might itself turn out to be the most stable.

\section{Fourth-order equations of motion}\label{sec:equationsOfMotion}

The equations of motion derived from the action \eqref{HDGaction} are
\bse\label{generalEOM}
\bea
H_{\mu\nu} & := &\, \dfrac{1}{\sqrt{-g}}\dfrac{\delta I}{\delta g^{\mu\nu}}  \label{eq:EOMgeneralMetric}\\
%H_{\mu\nu} 
& = &\, 
\gamma \left(R_{\mu\nu}-\half g_{\mu\nu}R\right)+\frac{2}{3}\left( \alpha-3\beta \right) \nabla_{\mu} \nabla_{\nu} R
-2\alpha \Box R_{\mu\nu}
+\frac{1}{3}\left( \alpha +6\beta \right)g_{\mu\nu} \Box R  \notag\\
& &
 -4\alpha R^{\eta\lambda}R_{\mu\eta\nu\lambda}
 +2\left( \beta +\frac{2}{3}\alpha \right) R R_{\mu\nu}
+\half g_{\mu\nu} \left(2 \alpha R^{\eta\lambda}R_{\eta\lambda}-\left( \beta +\frac{2}{3}\alpha \right) R^{2}\right) \notag\\
 &=&\half T_{\mu\nu}\,,  \label{eq:EOMgeneralMetricsourced}
\eea
\ese
satisfying a generalised Bianchi identity
\begin{equation}\label{eq:generalisedBianchi}
\nabla^{\nu}H_{\mu\nu} \equiv 0
\end{equation}
and with trace
\begin{equation}\label{eq:EOMtrace}
H_{\mu}^{\;\mu} = 6 \beta \Box R -\gamma  R = \half T_{\mu}^{\;\mu}\,,
\end{equation}
which is of fourth-order in derivatives of the metric for $\beta\ne0$ and of second-order for $\beta=0$.
In fact the $\beta=0$ (Einstein-Weyl) theory will turn out to be of particular interest to us. Note that in the Einstein-Weyl theory we can identify the more desirable sign of $\alpha$ by
linearising around a Minkowski background. Writing $g_{\mu\nu}=\eta_{\mu\nu}+
h_{\mu\nu}$, we find
\begin{equation}
-\ft13\alpha \Box (\Box - \frac{\gamma}{\alpha}) h_{\mu\nu} =0\,.
\end{equation}
And so $\alpha > 0$ is required for the absence of tachyonic instabilities.

From the study of the linearised limit of \eqref{generalEOM} about flat spacetime in \cite{Stelle:1977ry}, one knows that there are massive spin-two and spin-zero excitations with masses
\bse
\begin{eqnarray}
m_2^{\;2} & := &\, \frac{\gamma}{2 \alpha}\,,\label{m2val}\\
m_0^{\;2} & := &\, \frac{\gamma}{6 \beta}\,,\label{moval}
\end{eqnarray}
\ese
so one notes that
\begin{equation}
H_{\mu}^{\;\mu} = 6 \beta \left(\Box  -m_0^{\;2} \right)R\,.
\end{equation}

When considering spherically-symmetric static solutions, we may take the metric to have the Schwarzschild form
\begin{equation}
ds^{2}=-B(r)\, dt^{2}+A(r)\, dr^{2}+r^{2}d\theta^{2}+r^{2}\sin^{2}\theta\: d\phi^{2}\,,
\end{equation}
and we shall look for source-free solutions with $T_{\mu\nu}=0$ in the bulk of spacetime.\footnote{We shall address the issue of delta-function sources for solutions in Sections \ref{sec:weakFieldExpansion} and \ref{sec:nonlinshellsources}.}
With this ansatz there are only two independent equations of motion, for $A(r)$ and $B(r)$. The $H$ field equation tensor takes the form 
\begin{equation}
H_{\mu\nu}=\begin{pmatrix}
H_{tt}(r) 	& 0 			& 0 						& 0\\
0 			& H_{rr}(r)	& 0 						& 0\\
0 			& 0 			& H_{\theta\theta}(r)	& 0\\
0 			& 0 			& 0 						& H_{\theta\theta}(r) \sin^2\theta\\
\end{pmatrix}\,,
\end{equation}
the components of which are related (for $\theta=\frac{\pi}{2}$) by the $r$ component of the Bianchi identity (\ref{eq:generalisedBianchi}):
\begin{equation}\label{eq:generalisedBianchiThTh}
\left(\frac{H_{rr}}{A}\right)'+\frac{2H_{rr}}{Ar}+\frac{B'H_{rr}}{2AB}-\frac{2H_{\theta\theta}}{r^3}+\frac{B'H_{tt}}{2B^2} \equiv 0\,.
\end{equation}
Accordingly, when provided with a $T_{\mu\nu}$ stress-tensor source, the system is described by just two independent equations
\bse\label{Httrreqns}
\begin{eqnarray}
H_{tt} & = &\, \half T_{tt}\,, \label{Htteqn} \\
H_{rr} & = &\, \half T_{rr}\,. \label{Hrreqn}
\end{eqnarray}
\ese

If the metric (\ref{eq:metric}) is substituted into the Lagrangian before performing the variation, the resulting equations of motion are found to be equivalent to the set \eqref{Httrreqns}, \ie the truncation to the static spherically-symmetric case is a consistent truncation:
\bse
\begin{eqnarray}
\dIdA & = &\, -\frac{\sqrt{-g}}{A^2} H_{rr}\,, \\
\dIdB & = &\, -\frac{\sqrt{-g}}{B^2}H_{tt} \,.
\end{eqnarray}
\ese
Consistency of truncation to the metric form \eqref{eq:metric} is guaranteed in the usual fashion because one is truncating to the invariant sector under a group action -- in this case spatial rotations \cite{Deser:2003up}. It should be emphasised at this point that in this paper we are not making any additional simplifying truncations such as setting $AB=\text{constant}$. Imposing such additional conditions certainly makes solution of the equations greatly simpler, but it also severely restricts the corresponding solution set.\footnote{Indeed, in \cite{Deser:2003up}, among other cases, the pure $C_{\mu\nu\rho\sigma}C^{\mu\nu\rho\sigma}$ theory was considered subject to such a restricted ansatz, with the result that the only solution without conical singularities is just the classic Schwarzschild solution. The same simplifying restriction has been made recently in the analysis of spherically symmetric solutions for the pure $R^2$ theory in Ref.\ \cite{Kehagias:2015ata}. Our aim in the present paper is to explore the full set of spherically symmetric solutions without such a restriction.}

From here onwards, unless otherwise stated, we will be solving the source-free equations for $r>0$
\bse\label{vacEOM}
\begin{eqnarray}
H_{tt} & = &\, 0\,, \label{eq:EOMHtt}\\
H_{rr} & = &\, 0 \,.\label{eq:EOMHrr}
\end{eqnarray}
\ese

\subsection{Differential order}\label{sec:differentialOrder}

\subsubsection{$ \beta \neq 0$}

To find the differential order of these coupled equations, 
note that $H_{tt}$ is a function of $A^{(3)}(r),B^{(3)}(r),B^{(4)}(r)$ and 
lower-order derivatives,\footnote{We denote derivatives of order $\ge3$ by 
superscripts such as $A^{(3)}$.} and $H_{rr}$ is a function of 
$A''(r),B''(r),B^{(3)}(r)$ and lower-order derivatives. Let us now analyse 
the differential order of these equations. Note that 
$\alpha= 0$ and $\beta=0$ are special cases of 
different differential order, so we shall first look at the generic case 
$\alpha \neq 0$, $\alpha\neq3\beta$, $\beta\neq 0 $.
We define 
\bse\label{X&Y}
\begin{IEEEeqnarray}{ll}
X(r)=\ & \frac1{A^2 \left(r (\alpha -3 \beta ) B'-2 (\alpha +6 \beta ) B\right)^2}\notag\\
&\ \ \times\Big((\alpha -3 \beta ) B \left(2 r B A' \left(r (\alpha -3 \beta ) B'-2 (\alpha +6 \beta ) B\right)\right.\notag\\
&\ \ \ \ +A \left(-r^2 (\alpha -3 \beta ) B'^2-4 r (\alpha -3 \beta ) B B'+12 (\alpha +6 \beta ) B^2\right)\Big)\,,\\
Y(r)=\ &\frac{2 r (\alpha -3 \beta ) B^2}{A \left(2 (\alpha +6 \beta ) B-r (\alpha -3 \beta ) B'\right)}\,.
\end{IEEEeqnarray}
\ese

It is clear that the equations of motion are equivalent to the pair
\bse\label{eq:General}
\begin{eqnarray}
0&=& H_{rr}\,, \label{eq:GeneralA} \\
0&=& H_{tt} -X(r)H_{rr} -Y(r) \partial_r H_{rr}\,, \label{eq:GeneralBB4B3} 
\end{eqnarray}
\ese
the first of which is of third order in B and of second order in A, and the second of which is of third order in A and of second order in B. Details are given in Appendix \ref{app:fullequations}. This reduction in order leads us to expect a total of six free parameters in the solution. This can be more clearly seen by eliminating $B(r)$ to get an equation of sixth order in $A(r)$ alone; the detailed procedure is sketched in Appendix \ref{app:fullequations}.

\subsubsection{$ \beta = 0 $}\label{sec:betazero}

In this special case, the massive scalar is absent.
In the trace of the sourceless equations of motion (\ref{eq:EOMtrace}) with $T_{\mu\nu}=0$, two derivatives disappear and the equation simply states that $R=0$. As this suggests, the total differential order in this case is reduced by two with respect to the $\beta \neq 0$ case. 
The equations of motion are then equivalent to the pair
\bse
\begin{IEEEeqnarray}{rCl}
%--------------------
0 &= &\, H_\mu^{\;\mu} \,,
\\
%--------------------
0 &= &\, \frac{H_{rr}}{\alpha}
%--------------------
 + H_\mu^{\;\mu} 
 \frac{3 r B A'-2 A \left(r B'+B\right)+2 A^2 B}{3 \gamma  r^2 A B}
 %--------------------
-(H_\mu^{\;\mu})^2 
\frac{A}{6 \gamma ^2}
%--------------------
-\partial_r(H_\mu^{\;\mu}) 
\frac{ 2 B-r B' }{ 3 \gamma  r B }\,,\hspace{1cm}
%--------------------
\end{IEEEeqnarray}
\ese
the first of which is of second order in B and of first order in A, and the second of which is of second order in A and of first order in B.

These two second-order equations imply that there are four free parameters in the solution for the $\beta=0$ case.

\section{Solutions near the origin and Frobenius analysis}\label{sec:solutionsNearTheOrigin}
Previously in Ref.\ \cite{Stelle:1977ry}, the asymptotic behaviour of solutions to the equations of motion was analysed near the origin, working to leading orders in $r$. Here we will solve expansions to several higher orders in $r$ in order to improve our understanding of the parametric dependences of solutions. 

The two undetermined functions in the metric are expanded in Frobenius series in $r$ as
\begin{equation}
\begin{split}\label{Frobseries}
A(r) & = a_{s}r^{s}+a_{s+1}r^{s+1}+a_{s+2}r^{s+2}+\ldots \,,\\
B(r) & = b_{t} \left( r^{t}+b_{t+1}r^{t+1}+b_{t+2}r^{t+2}+\ldots \right) \,,
\end{split}
\end{equation}
where $a_{s},b_{t} \neq 0$ are nonvanishing coefficients and $s,t$ are indices yet to be determined.

Substituting the series \eqref{Frobseries} into the equations of motion \eqref{vacEOM} and analysing the consistent possibilities for the $(s,t)$ indices leads exclusively to three solution families \cite{Stelle:1977ry} for generic\,\footnote{We neglect here families of solutions appearing only at special values of $\alpha > 3 \beta > 0$:
\be
\frac{t-2}{3}  = s \in \mathbb{Z}^+\ ,\hspace{.5cm}
\alpha  = \frac{  (s^2+2s+2)^2}{s^4}3 \beta\notag
\ee
\indent and two families of solutions for $\alpha=0$:
\be
 \frac{4+2t+t^2}{4+t}  =  s \in \mathbb{Z}^+
 \hspace{.5cm}\mbox{or}\hspace{.5cm}
 \ (s,t)  = (0,1) \text{ (with 1 free parameter)}\,.\notag
\ee
} $\alpha\ne0$, $\beta\ne0$:
\bea
\bullet\quad (s,t) &=& (0,~~0)\notag\\
\bullet\quad (s,t) &=& (1,-1)\label{asympfamilies}\\
\bullet\quad (s,t) &=& (2,~~2)\notag
\eea
In each of these families, the equations can be solved order-by-order for the coefficients $a_{n},b_{n}$. Some coefficients will be left undetermined in this process, corresponding to the free parameters of the system in each solution family. There will always be one free parameter in $B(r)$ corresponding to a trivial scaling of the time coordinate.

Of course, in performing asymptotic analysis of this sort, an assumption is being made that Frobenius type expansions such as those of Eqs.\ \eqref{Frobseries} with integral steps in powers of $r$ following the leading $(r^s,r^t)$ terms is adequate to capture all possible types of asymptotic behaviour for solutions to nonlinear equations such as \eqref{vacEOM}. One might worry about the inclusion of terms such as $\exp(c/r^p)$ times a Frobenius series, or of terms involving powers of logarithms. In the case of linear systems of differential equations, one can deal with such possibilities on the basis of general theorems about equation systems with regular or irregular singular points of various ranks, but a suspicion could remain that this might not capture the full complexity of solutions to systems such as \eqref{vacEOM} (written out in full detail in Appendix \ref{app:fullequations}). All we can say to dispel such concerns is that we have explicitly tried many such exotic possibilities and the only consistent leading asymptotic behaviours that we have found are those shown in \eqref{asympfamilies}.

\subsection{Free-parameter counts in each of the near-origin solution families}

We have expanded and solved the equations of motion \eqref{vacEOM} to at least twelve orders in $r$. In each family, all free parameters have appeared by the fourth order at the latest, and after that each new order brings two new parameters and two new constraints. The resulting free-parameter counts are given in Table \ref{tab:solparametercounts}:
\begin{table}[ht]
\centering
\caption{Free parameter counts for the indicial solution families.\label{tab:solparametercounts}}
\begin{tabular}{|c|c|l|}
\hline
$(s,t)$ solution family & number of free parameters & a choice of free parameters \\
\hline
$(0,\;\; 0)$ & 3 & $b_0,a_2,b_2$\\
$(1,-1)$ & 4 & $a_1,b_{-1},a_4,b_2$\\
$(2,\;\; 2)$ & 6 & $a_2,b_2,b_3,b_4,a_5,b_5$\\
\hline
\end{tabular}
\end{table}

\subsubsection{The $(0,0)$ family}\label{sec:smallr00}
The first few terms in the $(0,0)$ family are:
\bse
\begin{IEEEeqnarray}{rl}\label{eq:generalAlphaBeta00A}
A(r) =\   &
1 + a_2 r^2
\notag\\ & 
+r^4 \frac{a_2 b_0 \left(b_0 \gamma  (2 \alpha +3 \beta )-36 \alpha  \beta  b_2\right)+18 a_2^2 \beta  b_0^2 (10 \alpha +3 \beta )-2 b_2 \left(b_0 \gamma  (\alpha -3 \beta )+9 \beta  b_2 (2 \alpha +3 \beta )\right)}{180 \alpha  \beta  b_0^2}
\notag \\ & 
+ O(r^6)\,,
\\ \label{eq:generalAlphaBeta00B}
\frac{B(r)}{b_0} =\  & 
1
+b_2 r^2
\notag\\ & 
+\frac{r^4 \left(54 a_2^2 \beta ^2+a_2 \left(-\alpha  \gamma +108 \alpha  \beta  b_2+3 \beta  \gamma \right)+b_2 \left(\gamma  (\alpha +6 \beta )+54 \beta  b_2 (2 \alpha -\beta )\right)\right)}{360 \alpha  \beta }
\notag \\ & 
+ O(r^6)\,.
\end{IEEEeqnarray}
\ese

The three-parameter $(0,0)$ solution is the natural `vacuum' solution family of the higher-derivative theory, comparable to the two-parameter spatially homogeneous flat space solution in Einstein theory. The Riemann curvature tensor $R_{\underline{abcd}}$ referred to an orthonormal frame is nonsingular as $r\to0$ for this solution.

\subsubsection{The $(1,-1)$ family}\label{sec:smallr1m1}
The first few terms in the $(1,-1)$ family are:
\bse
\begin{IEEEeqnarray}{ll}
\label{eq:generalAlphaBeta1m1A}
A(r) =\  &
a_1 r
-a_1^2 r^2
+a_1^3 r^3
+a_4 r^4
-\frac{1}{16} r^5 \left(a_1 \left(3 a_1 b_2+19 a_1^4+35 a_4\right)\right)
\notag\\ &  
+\frac{1}{40} a_1^2 r^6 \left(21 a_1 b_2+101 a_1^4+141 a_4\right)
+O(r^7)\,,
\\ \label{eq:generalAlphaBeta1m1B}
\frac{B(r)}{b_{-1}} =\ &
\frac{1}{r}
+a_1
+b_2 r^2
+\frac{1}{16} r^3 \left(a_1 b_2+a_1^4+a_4\right)
\notag\\  & 
-\frac{1}{40} 3 r^4 \left(a_1 \left(a_1 b_2+a_1^4+a_4\right)\right)
+O(r^5)\,.
\end{IEEEeqnarray}
\ese

The $(1,-1)$ family is clearly the family that contains the classic Schwarzschild solution of Einstein theory. The Schwarzschild solution is obviously a solution of the higher-derivative theory's field equations because every term in \eqref{generalEOM} contains $R_{\mu\nu}$ or $R$. At the origin, the $(1,-1)$ indicial structure gives rise to a curvature singularity, with $R_{\mu\nu\rho\sigma}R^{\mu\nu\rho\sigma}$ going like $r^{-6}$ as $r\to0$ \cite{Stelle:1977ry}.

\subsubsection{The $(2,2)$ family}\label{sec:smallr22}
The first few terms in the $(2,2)$ family are:
\bse
\begin{IEEEeqnarray}{ll}
A(r) =\  & 
a_2 r^2
+a_2 b_3 r^3
-\frac{a_2 r^4}{6} \left(2 a_2+b_3^2-8 b_4\right)
+a_5 r^5
\notag\\ & 
+\frac{r^6}{1296 \alpha  \beta } 
\Bigg(
-12 \alpha ^2 a_2^3-2 a_2^2 \left(
	b_3^2 \left(
		\alpha ^2-603 \alpha  \beta -252 \beta ^2\right)
	+27 \alpha  \left(
		20 \beta  b_4+\gamma \right)
	\right)
\notag\\ & 
+a_2 \Big(
	b_3^4 \left(
		-16 \alpha ^2+1413 \alpha  \beta -72 \beta ^2\right)
	+2 b_4 b_3^2 \left(
		19 \alpha ^2-2223 \alpha  \beta +180 \beta ^2\right)
	\notag\\ &
	-36 b_5 b_3 \left(
		\alpha ^2+45 \beta ^2\right)
	+12 \alpha  b_4^2 (\alpha +162 \beta )\Big)
+324 a_5 \beta  b_3 (7 \alpha +3 \beta )
\Bigg)
\notag\\ & 
+O(r^7)\,,
\\
\frac{B(r)}{b_2} =\  & 
r^2
+b_3 r^3
+b_4 r^4
+b_5 r^5
\notag\\ & 
+\frac{r^6 }{216 \alpha  a_2} 
\Bigg(
	-12 \alpha  a_2^3+a_2^2 \left(
		14 b_3^2 (2 \alpha +3 \beta )-24 \alpha  b_4\right)
	\notag\\ & 	
	+a_2 \left(
		2 b_3^4 (67 \alpha -3 \beta )+2 b_4 b_3^2 (15 \beta -227 \alpha )+45 b_5 b_3 (7 \alpha -3 \beta )+180 \alpha  b_4^2\right)
	+27 a_5 b_3 (\alpha +3 \beta )\Bigg)
\notag\\&
 +O(r^7)\,.
\end{IEEEeqnarray}
\ese

The $(2,2)$ indicial structure at the origin gives rise to a curvature singularity for this solution family, with $R_{\mu\nu\rho\sigma}R^{\mu\nu\rho\sigma}$ going like $r^{-8}$ as $r\to0$ \cite{Stelle:1977ry}.

\section{Coupling to sources in the linearised theory}\label{sec:weakFieldExpansion}

For asymptotically-flat solutions, the weakening fields as $r\to\infty$ can reliably be analysed using the linearised limit of the field equations \eqref{generalEOM}. We now set the stage for our later discussion of source coupling in the full nonlinear theory by studying  coupling  to sources in the linearised theory, expanding somewhat the discussion given in Ref.\ \cite{Stelle:1977ry}. We first need to identify the vacuum solutions that can occur inside a shell source.

\subsection{Source-free solutions}
In \cite{Stelle:1977ry}, the linearised equations were solved for
\bse\label{eqa:linmetric}
\begin{eqnarray}
A & = &\, 1+W(r)+O(W^2)\,, \\
B & = &\, 1+V(r)+O(V^2)\,.
\end{eqnarray}
\ese
Solving the linearised source-free equations \eqref{vacEOM} for $r>0$ in this limit yields the general solution
\bse\label{eqa:linearisedgensol}
\begin{eqnarray}
V & = &\, C+\frac{C_{2,0}}{r}+\frac{C_{0-} e^{-m_0 r}}{r}+\frac{C_{0+} e^{m_0 r}}{r}+\frac{C_{2-} e^{-m_2 r}}{r}+\frac{C_{2+} e^{m_2 r}}{r} \label{eq:linearisedV}
\\
W & = &\, -\frac{C_{2,0}}{r}+C_{0-} \frac{e^{-m_0 r}}{r}\left(1+m_0 r \right)+C_{0+} \frac{e^{m_0 r}}{r}\left(1-m_0 r\right) \notag
\\ &&
 -\half C_{2-}  \frac{e^{-m_2 r}}{r}\left(1+m_2 r\right)-\half C_{2+} \frac{e^{m_2 r}}{r}\left(1-m_2 r\right)\,.\label{eq:linearisedW}
\end{eqnarray}
\ese
Note here that for $\alpha < 0$ or $\beta < 0$ one has pure imaginary $m_2$ or pure imaginary $m_0$, respectively. For pure imaginary masses $m_i=i\mu_i$ the solution has, instead of real exponentials, suppressed oscillating terms like $C_{i,s}\,\frac{1}{r}\sin(\mu_i r)$ and $C_{i,c}\,\frac{1}{r}\cos(\mu_i r)$ in $V$ and $W$. However, $W$ also has non-suppressed oscillations behaving like $C_{i,s}\sin(\mu_i r)$ and $C_{i,c}\cos(\mu_i r)$. This precludes asymptotic flatness at spatial infinity unless both of the constants $C_i$ vanish, \ie unless the corresponding metric solution is strictly flat. Accordingly, we limit our consideration to cases where $m_2\ge0$ and $m_0\ge0$.

The linearised solution \eqref{eqa:linearisedgensol} clearly shows the existence in general of six free parameters, noting that the free parameter $C$ corresponds to a trivial rescaling of the time coordinate. As one approaches the origin in the linearised solution \eqref{eqa:linearisedgensol} for generic values of the free parameters, the Cartesian-coordinate form of the linearised curvature tensor $R^{\text{lin}}_{\underline{abcd}}$ has leading $r^{-3}$ singular terms in the $r\to0$ limit. The linearised Ricci scalar for the solution \eqref{eqa:linearisedgensol} is
\be
R = -\frac{3 C_{0-} m_0^2 e^{- m_0 r}}{r}-\frac{3 C_{0+} m_0^2 e^{m_0 r}}{r}\label{eq:linricci}
\ee
and so has leading $r^{-1}$ behaviour for generic values of the free parameters.

Note that $\alpha=0$ or $\beta=0$ or $\alpha=3\beta$ are special cases in which $m_2$ or $m_0$ vanish or coincide. In the following, we shall proceed for the generic case $\alpha \neq 0$, $\beta \neq 0$, $\alpha \neq 3\beta$.

\subsection{True linearised vacuum}

When the general linearised solution \eqref{eqa:linearisedgensol} is extended all the way to the origin at $r=0$, $H_{\mu\nu}$ must in general involve $\delta^3(\vec{r})$ sources. The true vacuum solution without such delta-function sources is the restricted three-parameter solution family satisfying the vacuum constraints 
\be
C_{2,0} = C_{2-}+C_{2+} = C_{0-}+C_{0+}=0\,.\label{eq:vacconstraints}
\ee
Subject to these constraints, one finds the true linearised vacuum family
\bse\label{eq:linearisedPureVacuum}
\begin{eqnarray}
V_{\text{vac}} & = &\, C+C_{0+}\frac{2 \sinh \left(m_0 r\right)}{r}+C_{2+}\frac{2 \sinh \left(m_2 r\right)}{r}\\
W_{\text{vac}} & = &\,   2 C_{0+}\left(\frac{\sinh(m_0 r)}{r} - m_0 \cosh(m_0 r) \right) 
        -   C_{2+}\left(\frac{\sinh(m_2 r)}{r} - m_2 \cosh(m_2 r) \right)\,.\notag\\
\end{eqnarray}
\ese
A consequence of the vacuum constraints \eqref{eq:vacconstraints} is that the metric \eqref{eqa:linmetric} for the true linearised vacuum family \eqref{eq:linearisedPureVacuum} is nonsingular at the $r=0$ origin. This nonsingularity extends as well to all components of the linearised curvature tensor $R^{\text{lin}}_{\underline{abcd}}$ and in particular one can see from \eqref{eq:linricci} using \eqref{eq:vacconstraints} that the linearised Ricci scalar is nonsingular at the origin for the true vacuum solution \eqref{eq:linearisedPureVacuum}.

It is appropriate to distinguish the true vacuum solution \eqref{eq:linearisedPureVacuum}, with its nonsingular curvature, from other members of the general solution family \eqref{eqa:linearisedgensol} that happen to have a nonsingular metric as $r\to0$. Unlike the situation in linearised Einstein theory, where the only spherically symmetric solution with a nonsingular metric at $r=0$ is simply flat space, with correspondingly vanishing curvature, in the linearised version of the higher-derivative theory \eqref{HDGaction} the family of nonsingular-metric solutions turns out to be wider than just the vacuum solution \eqref{eq:linearisedPureVacuum}. This wider class of nonsingular-metric solutions includes also the solution for a point delta-function source, which we consider next.

\subsection{Source examples for the linearised theory}
\subsubsection{Point source}\label{ssec:pointsource}
In \cite{Stelle:1977ry} the stress-tensor of a static point mass at the origin was considered:
\begin{equation}
T_{\mu\nu}=\delta_{\mu}^0 \delta_{\nu}^0 M \delta^3(\vec{x})\,.\label{eq:deltasource}
\end{equation}
With this source, the solution to the linearised equations of motion is the vacuum solution plus an asymptotically-flat matter part: 
\bse\label{eqa:linearisedDeltaSourceSol}
\begin{eqnarray}
V(r) & = &\, C -\frac{M}{24 \pi  \gamma r} \left(e^{-m_0 r}-4 e^{-m_2 r}+3\right)\,, \label{Vpoint}\\
W(r) & = &\, -\frac{M }{24 \pi  \gamma r}\left(\left(1+m_0 r\right)e^{-m_0 r} +2\left(1+m_2 r\right) e^{-m_2 r} -3\right)\,,
\end{eqnarray}
\ese
indicating that one needs $\gamma = {1\over16\pi G}={2\over\kappa^2}$ in order to agree with the Schwarzschild solution in the limit where $m_0$ and $m_2$ tend to infinity.

As one can see from \eqref{eqa:linearisedDeltaSourceSol}, $V$ and $W$ are actually nonsingular as $r\to0$. As is clear from the need for the source \eqref{eq:deltasource}, however, \eqref{eqa:linearisedDeltaSourceSol} cannot be considered a true vacuum solution. This conclusion is reinforced by consideration of the curvature $R^{\text{lin}}_{\underline{abcd}}$ as $r\to0$, whose components have leading $r^{-1}$ singularity, and for which the Ricci scalar is given by
\be
\label{eq:linearisedDeltaSourceRicci}
R = \frac{ M }{8 \pi  \gamma r} \; m_0^2 \; e^{-m_0 r}\,,
\ee
which has $r^{-1}$ behaviour as $r\to0$. Note also that in the weak-field regime with $m_0$ finite (\ie for $\beta \neq 0$) a solution with a point source at $r=0$ always has $R \neq 0$ at any nonzero value of $r$.

\subsubsection{Shell source}\label{sec:shellSourcesLinearisation}
To illustrate the effects of extended sources in this theory with no Birkhoff theorem, let us now solve again in the linearised theory for the fields produced by various sources of nonzero size.

First take as source a thin spherical shell of radius $\ell$
\bse\label{shellSource}
\begin{eqnarray}\label{eq:shellSourceDef}
T_{tt} & = &\, \frac{M}{4 \pi \ell^2}\delta(r-\ell)\,, \\
T_{rr} & = &\, 0\,.
\end{eqnarray}
\ese
From the linearised $\nabla^{\mu}T_{\mu\nu} = 0$ condition, we have
\begin{equation}
T_{\theta\theta} = 0 + O((W,V)^2)\,.
\end{equation}
For $r<\ell$ we use the vacuum solution (\ref{eq:linearisedPureVacuum}):
\bse\label{linvacsol}
\begin{eqnarray}\label{eq:linearisedPureVacuumD}
V_{\text {in}} & = &\, D - \frac{2 D_{0-}\sinh(m_0 r)}{r} - \frac{2 D_{2-}\sinh(m_2 r)}{r}\,, \\
W_{\text {in}} & = &\,   - 2 D_{0-}\left(\frac{\sinh(m_0 r)}{r} - m_0 \cosh(m_0 r) \right) 
        +   D_{2-}\left(\frac{\sinh(m_2 r)}{r} - m_2 \cosh(m_2 r) \right)\,,\notag\\
\end{eqnarray}\ese
and for $r>\ell$ we use the source-free solution \eqref{eq:linearisedV} for $V_{\text {out}}$ and \eqref{eq:linearisedW} for $W_{\text {out}}$ with the rising exponentials suppressed in order to achieve asymptotic flatness.

For $\alpha\neq 0$ and $\beta \neq 0$, one finds that $V'''(r)$ and $W''(r)$ are discontinuous at the location of the shell, and the solution is
\bse\label{linextsol}
\begin{eqnarray}\label{eq:linearisedShellSourceSol}
V_{\text{out}} & = &\, 
D
+\frac{M}{8 \pi \gamma }\left(\frac{1}{\ell}-\frac{1}{r}\right)
+\frac{e^{-m_2 r}}{r}\frac{M  }{ 6 \pi  \gamma  }\frac{\sinh (m_2 \ell)}{m_2 \ell}
-\frac{e^{-m_0 r}}{r}\frac{M  }{24 \pi  \gamma  }\frac{\sinh (m_0 \ell)}{m_0 \ell}\,,\label{Vshell}
\\
W_{\text{out}} & = &\, -\frac{M  }{24 \pi  \gamma r}\left(
\left(1+m_0 r\right)e^{-m_0 r}\frac{\sinh (m_0 \ell)}{m_0 \ell}
+2\left(1+m_2 r\right)e^{-m_2 r}\frac{\sinh (m_2 \ell)}{m_2 \ell}
-3
\right)\,,\notag\\
\end{eqnarray}\ese
for which the Ricci scalar is
\begin{equation} \label{eq:linearisedShellSourceRicci}
R_{\text{out}} = \frac{ M }{8 \pi  \gamma r} \; m_0^2 \; e^{-m_0 r} \; \frac{\sinh(m_0 \ell)}{m_0 \ell}\,.
\end{equation}
Thus, just as for the point source, we find in the weak-field regime with $m_0$ finite (\ie for $\beta \neq 0$) that the solution with a shell source always has $R \neq 0$ at any nonzero value of $r$.

Note that in the limit $\ell \to 0$ of the shell-source solution, the expressions for $V_{\text {out}}$, $W_{\text {out}}$ and $R$ correctly tend to those of the point source.

\subsubsection{Balloon source}\label{sec:balloonsource}
Now let us take as source a stress tensor with internal pressure, again expanding upon results given in \cite{Stelle:1977ry}:
\begin{equation}
T_{\mu\nu}=\begin{pmatrix}
\frac{3 M}{4 \pi \ell^3}\Theta(\ell-r)	 & 0 			& 0 						& 0\\
0 		& 		P \Theta(\ell-r)		& 0 						& 0\\
0 		& 		0					& T_{\theta\theta} 						& 0\\
0 		& 		0					& 0 						& T_{\theta\theta} \sin^2\theta\\
\end{pmatrix}\,,
\end{equation}
where $\Theta(r)$ is a Heaviside theta function.\footnote{$\Theta(r)$ is capitalised in order not to be confused with the angular coordinate $\theta$.}
In order to satisfy the linearised conservation condition for $T_{\mu\nu}$, we need to have
\begin{equation}
T_{\theta \theta } = P r^2 \Theta (l-r)-\frac{1}{2} P r^3 \delta (l-r)\,.
\end{equation}
Solving the system with this source, we find the interior $r < \ell$ solution
\bse
\begin{IEEEeqnarray}{ll}
V_{\text{in}}(r) =\ & -\frac{2 D_{0-} \sinh \left(m_0 r\right)}{r}-\frac{2 D_{2-} \sinh \left(m_2 r\right)}{r}+D+\frac{r^2 \left(4 \pi  l^3 P+M\right)}{16 \pi  \gamma  \ell^3}\,,\\
W_{\text{in}}(r) =\ & D_{0-} \left(2 m_0 \cosh \left(m_0 r\right)-\frac{2 \sinh \left(m_0 r\right)}{r}\right)\notag\\
&\ \ +D_{2-} \left(\frac{\sinh \left(m_2 r\right)}{r}-m_2 \cosh \left(m_2 r\right)\right)+\frac{M r^2}{8 \pi  \gamma  \ell^3}\,,
\end{IEEEeqnarray}\ese
where the three vacuum constraints $0 = D_{2,0} = D_{2-}+D_{2+} = D_{0-}+D_{0+}$ have again been used to ensure a pure-vacuum $r < \ell$ internal solution without source. (Note that the $M,P$ source terms are proportional to $r^2$ and do not affect these constraint requirements.)
At $r=l$ there are 5 continuity conditions, for $V,V',V'',W,W'$, and two step conditions
\bse
\begin{align}
V_{\text{out}}'''(\ell_+) = &\, V_{\text{in}}'''(\ell_-) + \frac{\ell P (\alpha +6 \beta )}{36 \alpha  \beta }
\\
W_{\text{out}}''(\ell_+)  = &\, W_{\text{in}}''(\ell_-)  - \frac{\ell^2 P (\alpha -3 \beta )}{36 \alpha  \beta }\,.
\end{align}\ese
Note, however, that of these seven continuity and step conditions, \emph{only six are independent} (\cf the six free parameters expected from the differential order analysis in Section \ref{sec:differentialOrder}).
This is a general situation, and will be important for us when we consider such source couplings in the full nonlinear theory.

Implementing the continuity and step conditions, we obtain the asymptotically flat solution with a balloon source:
\bse\label{eqa:balloonsource}
\begin{align}
V_{\text{in}} = & 
D
+ \frac{1}{48 \pi \gamma l} \Bigg(
3 M_P \frac{r^2}{l^2}
+2 \left[ 3 \, (1+m_0 l)M_0 - 4 \pi l^3 P \right] \frac{\sinh(m_0 r)}{m_0 r} e^{- m_0 l}
\notag\\&\qquad\qquad\quad
-8 \left[ 3 \, (1+m_2 l)\, M_2 + 2 \pi l^3 P \right] \frac{\sinh(m_2 r)}{m_2 r} e^{- m_2 l}
\Bigg)
\\
W_{\text{in}} = & 
\frac{1}{24 \pi \gamma l} \Bigg(
3 M \frac{r^2}{l^2}
+ \left[ 3 \, (1+m_0 l)\, M_0 - 4 \pi l^3 P \right] 
\left[ \frac{\sinh(m_0 r)}{m_0 r} - \cosh(m_0 r) \right] e^{- m_0 l}
\notag\\&\qquad
+2 \left[ 3 \, (1+m_2 l)\, M_2 + 2 \pi l^3 P \right] 
\left[ \frac{\sinh(m_2 r)}{m_2 r} - \cosh(m_2 r) \right] e^{- m_2 l}
\Bigg)
\\
V_{\text{out}} = & 
D + \frac{1}{16 \pi \gamma l}\left( 2 M_0 - 8 M_2 + 3M + 4 \pi l^3 P \right)
- \frac{M}{8 \pi \gamma r}
\notag\\& 
+\frac{e^{-m_0 r}}{24 \pi \gamma r} \left( 3 M_0 \left[ \frac{\sinh(m_0 l)}{m_0 l} - \cosh(m_0 l) \right] -4 \pi l^3 P \frac{\sinh(m_0 l)}{m_0 l} \right)
\notag\\&
-\frac{e^{-m_2 r}}{6 \pi \gamma r} \left( 3 M_2 \left[ \frac{\sinh(m_2 l)}{m_2 l} - \cosh(m_2 l) \right] +2 \pi l^3 P \frac{\sinh(m_2 l)}{m_2 l} \right)
\\
W_{\text{out}} = & 
\frac{M}{8 \pi \gamma r}
\notag\\& 
+\frac{e^{-m_0 r}(1+m_0 r)}{24 \pi \gamma r} \left( 3 M_0 \left[ \frac{\sinh(m_0 l)}{m_0 l} - \cosh(m_0 l) \right] -4 \pi l^3 P \frac{\sinh(m_0 l)}{m_0 l} \right)
\notag\\&
+\frac{e^{-m_2 r}(1+m_2 r)}{12 \pi \gamma r} \left( 3 M_2 \left[ \frac{\sinh(m_2 l)}{m_2 l} - \cosh(m_2 l) \right] +2 \pi l^3 P \frac{\sinh(m_2 l)}{m_2 l} \right)
\end{align}
\ese
where we have used the following notation for source-parameter combinations
\bea
M_P &:=&  M + 4 \pi \l^3 P\notag
\\
M_0 &:=& \frac{M - 4 \pi l^3 P}{l^2 m_0^2}
\\
M_2 &:=& \frac{M + 2 \pi l^3 P}{l^2 m_2^2}\,.\notag
\eea

The main point to take away from this analysis of the linearised solutions is that the general six-parameter solution, constrained by two requirements of vanishing coefficients for the rising-exponential Yukawa terms as $r\to\infty$, has a remaining essential dependence on four parameters. One of these is adjustable by rescaling of the time coordinate $t$ (corresponding to the additive parameter $D$ above), and will be fixed by the requirement of having an asymptotic Minkowski metric as $r\to\infty$. The other three parameters will be fixed by details of the source, as displayed in the balloon-source solution by the dependence on $\ell$, $M$ and $P$. This multi-parameter dependence clearly illustrates the absence of a Birkhoff theorem for the higher-derivative gravity theory. One needs to start with the full six-parameter generic solution in order to arrange a successful coupling of the higher-derivative theory to a standard matter source, exemplified here by these various delta-function source constructions.

\section{Shell sources in the full nonlinear theory}\label{sec:nonlinshellsources}

We now progress to studying matter coupling in the full nonlinear theory. Unlike the situation in general relativity, where the Schwarzschild solution is known in closed form, we have no such luxury in the higher-derivative gravity theory. So we need to be careful in handling the continuity and step matching conditions for solutions known only from series expansions such as those given in Section \ref{sec:solutionsNearTheOrigin}. What we wish to establish is which of the three families $(s,t)=(0,0),\ (1,-1)\ \text{or}\ (2,2)$ can couple acceptably to an ordinary matter stress-tensor source. The key to this will be the parameter counts that we found in Section \ref{sec:solutionsNearTheOrigin}. 

For simple models of matter coupling, we again consider distributional sources. As has long been clear \cite{Geroch:1987qn} in general relativity, however, the only sensible delta-function sources in generally covariant theories are sources of spatial codimension one. So we do not consider a point source as in the linearised theory. Instead, the simplest source that we can consider in the full nonlinear theory is a thin spherical shell of radius $\ell$, which can be compared to the discussion given for the linearised theory in Section \ref{sec:shellSourcesLinearisation}. This shell source has a conserved stress tensor
\begin{equation}\label{eq:theSource}
T_{\mu\nu}=\begin{pmatrix}
T_{tt}	& 0 	& 0 				& 0\\
0 		& T_{rr}& 0 				& 0\\
0 		& 0 	& T_{\theta\theta}	& 0\\
0 		& 0 	& 0 				& T_{\theta\theta}\sin^2(\theta)\\
\end{pmatrix}\,,
\end{equation}
where, as in \eqref{shellSource}, 
\bse
\begin{align}
T_{tt} = &\, \frac{M}{4 \pi \ell^2}\delta(r-\ell)\,, \\
T_{rr} = &\, 0 \,.
\end{align}
\ese

The condition $\nabla^{\mu}T_{\mu\nu} = (0, \nabla^{\mu}T_{\mu r}, 0, 0) = 0$ requires
\begin{equation}\label{TthetaTtrel}
T_{\theta\theta} = \frac{r^3 B' T_{tt}}{4 B^2}\,.
\end{equation}

The equations of motion \eqref{generalEOM} expand schematically as
\bse
\begin{align}
H_{tt} = &\, \sim B^{(4)} + \sim A^{(3)} + \sim B^{(3)} + \ldots \,, \\
H_{rr} = &\, \sim B^{(3)} + \sim A'' + \sim B'' + \ldots \,,\\
H_{\theta\theta} = &\, \sim B^{(4)} + \sim A^{(3)} + \sim B^{(3)} + \ldots \,,
\end{align}\ese
suggesting that we should consider
\bse
\begin{eqnarray}\label{eq:stepMatchingsShell}
B^{(4)} & \sim & \delta + \Theta \,,\\
A^{(3)} & \sim & \delta + \Theta \,,\\
B^{(3)} & \sim & \Theta \,,\\
A''     & \sim & \Theta\,.
\end{eqnarray}\ese
Then $A,A',B,B',B''$ will be continuous at $r=\ell$, while $A''$ has a step of size
\begin{equation}
A_{\text{out}}''(\ell_+) - A_{\text{in}}''(\ell_-) =
\left. \frac{M}{8\pi \ell}A^3\frac{\ell(\alpha-3\beta)B'-2(\alpha+6\beta)B}{36 \alpha \beta} \right|_{r=\ell}\,.
\end{equation}

We leave to Appendix \ref{app:shellcoupling} a detailed discussion of how to arrange a satisfactory series solution of these matching conditions in the higher-derivative theory. 
However, the important part of the result is easily seen by a parameter-counting argument as follows.
The region interior to the shell is described by the vacuum solution of the nonlinear theory, which is the $(0,0)$ family as discussed in Section \ref{sec:smallr00}. The $(0,0)$ vacuum solutions of the nonlinear theory
admit three free parameters, as shown in Table \ref{tab:solparametercounts} given in Section \ref{sec:solutionsNearTheOrigin}.
The coupling to the source constitutes six continuity and matching conditions.
The region exterior to the shell is also source-free, so will be described by one of the source-free solutions of Section \ref{sec:solutionsNearTheOrigin}. Finally there are also two conditions at infinity that need to be imposed in order to ensure asymptotic flatness, analogous to the elimination of the rising-exponential Yukawa terms in the linearised theory.
After applying all these constraints we expect one final free parameter to be the adjustable parameter in the $B$ function, corresponding to the asymptotic value of $g_{00}$, which needs to equal $-1$ in order to have asymptotic Minkowski space as $r\to\infty$. 
The remaining structure of the solution will be determined by the details of the source, determined by the two parameters $M$ and $\ell$ in the case of the simple shell delta-function source, and determined by the three parameters $M$, $\ell$ and $P$ in the case of a balloon-type source as in Section \ref{sec:balloonsource}.
This is a total of 6+2+1=9 conditions for asymptotically Minkowski shell-coupled solutions, with 3 free parameters from the $(0,0)$ solution inside the shell, so the solution outside the shell must have 6 (or more) free parameters in order to be able to satisfy the constraints. Thus the exterior solution to a matter shell must be of the $(2,2)$ family, which has precisely 6 free parameters.
This is similar to the linearised source couplings where the exterior solutions had all of their 6 free parameters fixed by the parameters of the source, \ie the generic 6-parameter solution can be placed outside the source and can satisfy the necessary constraints, but a constrained exterior solution could not. Note that we assume that the 9 continuity, matching and asymptotic flatness conditions are all independent. In the linearised theory, one can verify that the conditions are indeed independent, but strictly speaking only a closed-form solution could confirm that the same is true in the full nonlinear theory.

On the other hand, trying to arrange the coupling of a delta-function shell source to an exterior $(0,0)$ or $(1,-1)$ family solution will not work, for the simple reason that their numbers of exterior free parameters (three or four, respectively) are not sufficient to satisfy all the nine continuity, step, asymptotic flatness and asymptotic Minkowskian requirements. Thus, an asymptotically-flat and asymptotically-Minkowskian solution coupled to a shell delta-function source can only be of the $(2,2)$ family.

This contrasts with general relativity, where the $(0,0)$ vacuum family has 1 free parameter, the $(1,-1)$ family has two free parameters, and the $(2,2)$ family does not exist. The shell coupling gives two conditions, $g_{00}(r \to \infty)=-1$ gives one condition, and there are no conditions needed to remove asymptotically non-flat terms. The solution exterior to a shell source is the Schwarzschild $(1,-1)$ family, with structure determined by $M$ and the asymptotic Minkowskian condition, but is independent of $\ell$.

\section{Trace-equation no-hair theorem}\label{sec:tracenohair}

Having established that the solution that couples correctly to an ordinary stress-tensor source is of the $(2,2)$ family, we now proceed to investigate the consequences of the field equations in the $(0,0)$, $(1,-1)$ and $(2,2)$ solution families without regard to sources.
We will be particularly interested in the consequences of boundary conditions at a putative horizon or at spatial infinity. 
Useful tools to this end are a set of Lichnerowicz-type 
`no-hair' theorems forcing the solution to share properties with the standard Schwarzschild solution under certain conditions. 
This topic was broached in Ref.\ \cite{Nelson:2010ig}. 
As noted in Ref.\ \cite{Lu:2015cqa}, we agree in part with conclusions of that reference, namely the trace part of the no-hair theorem, as will be discussed in the following. Unfortunately, we do not agree with the contentions of \cite{Nelson:2010ig} regarding the traceless part the higher-derivative equations of motion, which would have significantly simplified the analysis.
We present in Appendix \ref{app:Nelson} our analysis of the general no-hair theorem, including an extension to include a cosmological constant. Despite our disagreement with Ref.\ \cite{Nelson:2010ig}, we still can obtain important constraints on the solution families of the higher-derivative theory \eqref{HDGaction} using no-hair type arguments. 
In this section, we review the trace no-hair theorem of \cite{Nelson:2010ig}. In Section \ref{sec:horizonexp}, a careful analysis of the parametric structure of solutions containing a horizon will be given, and in Section \ref{sec:Schlinnohair} these elements will be put together with a no-hair theorem for {\it linearised} deviations from the classic Schwarzschild solution. In Section \ref{sec:numanalysis}, solutions with a horizon that discretely differ from the Schwarzschild solution will be discussed.
The only family of static spherically-symmetric and asymptotically-flat solutions that couples properly to ordinary stress-tensor sources, \ie the $(2,2)$ family, cannot have a horizon.

We now proceed to review the trace-equation no-hair theorem. We do this in a different style from that of Ref.\ \cite{Nelson:2010ig} in that we present the no-hair trace-equation argument for static solutions in terms of a timelike dimensional reduction from four to three dimensions.\footnote{Such timelike dimensional reductions have proven to be powerful tools in classifying black-hole solutions in a variety of supersymmetric and non-supersymmetric contexts \cite{blackholeclassification}.}

The static four-dimensional metric can be written in the 
form
%%%%%
\begin{equation}
ds^{2} = -\lambda^2 \, dt^{2}+h_{ab} \, dx^a dx^b\,,
\end{equation}
%%%%%
where the spatial metric $h_{ab}$ is positive definite for flat space, and 
therefore, assuming asymptotic flatness, it is positive definite 
everywhere between infinity and a horizon at finite $r$ (should one exist).
Both $h_{ab}$ and $\lambda$ are assumed to be functions only of the
three spatial coordinates $x^a$.
Let $\nabla_{\mu}$ be the covariant derivative for the 4-metric 
$g_{\mu\nu}$, and let $D_{a}$ be the covariant derivative for the 3-metric 
$h_{ab}$.  It follows that
%%%%%
\begin{equation}
\Box R := g^{\mu\nu}\nabla_{\mu}\nabla_{\nu} R = 
D^{a}D_{a} R + 
  \frac{1}{\lambda }\left(D^{a}\lambda \right)\left(D_{a}R\right)\,,
\end{equation}
%%%%%
so the trace of the source-free equations of motion (\ref{eq:EOMtrace})
can be written as
%%%%%
\begin{equation}\label{eq:EOMtraceDimReduced}
0 = H_{\mu}^{\;\mu} = 6\beta \left(
D^{a}D_{a} R + \frac{1}{\lambda }\left(D^{a}\lambda \right)
\left(D_{a}R\right) -m_0^{\;2} R \right)\,.
\end{equation}
%%%%%

   For the theory with $\beta = 0$, one has $R=0$ directly from the trace of the field equations \eqref{eq:EOMtrace}, 
while for $\beta\neq 0$ the bracketed quantity in (\ref{eq:EOMtraceDimReduced})
is required to vanish. 
Subject to certain assumptions, this will still imply that $R=0$. To see this, 
multiply the bracketed quantity in (\ref{eq:EOMtraceDimReduced}) by $\lambda\,R$ 
and integrate over the whole spatial 3-section
%%%%%
\begin{equation}
0=\int_{\cal S} \sqrt{h}\; {\rm d}^3x\left[
 \lambda\, R\; \left(D^a D_a R\right) + \lambda\, 
R \left( D^a \lambda  \right)\left( D_a R\right) - m_0^{\;2} \;\lambda\, R^2 
\right]\,.
\end{equation}
%%%%%
Preparing for an integration by parts, we rewrite this as
%%%%%
\begin{equation}\label{eq:NelsonThm2}
0=\int_{\cal S} \sqrt{h}\; {\rm d}^3x\left[
 D^a \left(\lambda\, R\; D_a R\right) -\lambda\,
\left( D^a R \right)\left( D_a R\right) -  m_0^{\;2} \; \lambda\, R^2 
\right]\,.
\end{equation}
%%%%%
The first term turns into a 2-dimensional integral over two boundaries:
one at spatial infinity, and the other at some finite radius. 
The contribution from the boundary  
at spatial infinity vanishes subject to the assumption of asymptotic flatness.
A sufficient (though not necessary) condition for the inner boundary 
term to vanish is satisfied if that boundary is a horizon.
If we take the inner integration boundary to be  such a horizon, 
then the boundary integral will be proportional to $\lambda|_{\text{horizon}}$,
which vanishes, by definition. Therefore, for a solution with a horizon, \eqref{eq:NelsonThm2}
reduces to a 3-dimensional spatial integral over the sum of two 
negative semi-definite terms.\footnote{Recall that we are requiring throughout $\beta>0$, so $m_0^2>0$.} Requiring this to vanish therefore implies that $R$ and $D_a R$
separately vanish throughout the integration region, thus
implying that $R=0$.

   In a region where $R=0$, the equations of motion become
%%%%%
\begin{equation}\label{eq:EOMwithR0}
0 = \left. H_{\mu\nu} \right|_{R=0} = 
-2 \alpha \left( 
\Box R_{\mu\nu}
+2 R_{\mu}^{\rho}R_{\nu\rho} - 2 \nabla_{\rho}\nabla_{\mu}R_{\nu}^{\rho}
-\frac{1}{2}g_{\mu\nu}R^{\rho\sigma}R_{\rho\sigma}
\right)
+\gamma R_{\mu\nu}\,,
\end{equation}
%%%%%
which notably no longer have any dependence on $\beta$. 
In the $\beta=0$ case without sources, one automatically has $H_{\mu}^{\;\mu}  = \gamma R=0$, 
and therefore (\ref{eq:EOMwithR0}) obtains everywhere. 

We have accordingly shown using  \eqref{eq:NelsonThm2} that if the boundary 
terms vanish on the boundaries of a given spatial region, then the field equations in that region reduce to 
the $\beta=0$ case (corresponding to the Lagrangian density 
${\cal L} \sim \gamma R -\alpha C_{\mu\nu\rho\sigma}C^{\mu\nu\rho\sigma}$). 
It will be computationally advantageous in such situations to use the two independent $\beta=0$ equations of motion, 
rather than the two $\beta\neq 0$ equations of motion together with $R=0$ as a 
third condition.

\subsection{Implications for the three near-origin solution families}
We have anticipated above a main conclusion that the asymptotically-flat solutions to the higher-derivative theory \eqref{HDGaction} with normal matter coupling do not have a horizon. For a spacetime without horizon it is natural to extend the integration region in Equation (\ref{eq:NelsonThm2}) down to near the origin, $r \to 0$, where $h_{ab}$ remains positive but we do not yet know the behaviour of the boundary term at the inner boundary. 
An analysis of the equations of motion at small $r$ can, however, tell us how this boundary term behaves.

We now look again at \eqref{eq:NelsonThm2} but now with the integration region taken to have an inner boundary located at $r \to 0$. We can use this discussion together with the small $r$ behaviour of the boundary term to study the various implications of \eqref{eq:NelsonThm2} for the structure of the $(0,0)$, $(1,-1)$ and $(2,2)$ solution families. Accordingly, we calculate the boundary term in (\ref{eq:NelsonThm2}) simply taken as the dominant part
\begin{equation*}
R\partial^rR 
\end{equation*}
(where the other components of the boundary term vanish in all cases as $r\to 0$). To be precise about the boundary-term contribution, recall that the boundary term actually appears in (\ref{eq:NelsonThm2}) in the form 
\begin{equation}
\int_{\cal S} \sqrt{h}\; \left[
 D_a \left( \sqrt{B}\; (\text{boundary term})^a\right)\right]{\rm d}^3x
\end{equation}
which for a boundary term with vanishing $\theta$,$\phi$ components is equal to
\begin{equation}
\int_{\cal S}\partial_r  \left[\sqrt{h}\;  \sqrt{B}\left(\text{boundary term}\right)^r\right]{\rm d}^3x =
\int_{\cal S}\partial_r  \left[\sqrt{AB}\; r^2 \sin(\theta)\left(\text{boundary term}\right)^r\right]{\rm d}r{\rm d}\theta{\rm d}\phi \\
\end{equation}
We now analyse the consequences of \eqref{eq:NelsonThm2} in the three near-origin solution families.

\vspace{10pt}
\begin{center}
\begin{tabular}{c|c|c}
Family & $R\partial^rR $ & $\sqrt{h \; B} \; R\partial^rR $
\\
\hlinevspaceThreeCol
$(0, 0)$ & $\frac{2 \gamma}{\beta} \left(a_2-b_2\right)^2 r + O(r^3)$ & $\sim O(r^3)$
\\[1.8ex]
\hlinevspaceThreeCol
$(1,-1)$ & $-\frac{3 \gamma}{8 a_1^4 \beta } \left(\frac{5}{3} a_1 b_2 - a_1^4 - a_4\right)^2 r + O(r^3)$ & $\sim O(r^3)$
\\[1.8ex]
\hlinevspaceThreeCol
$(2, 2)$ & $-\frac{\left(a_2 \left(14 a_2 b_3-2 b_3^3+10 b_4 b_3-45 b_5\right)+27 a_5\right){}^2}{9 a_2^5 }r^{-5} + O(r^{-4})$ & $\sim O(r^{-1})$
\\[1.8ex]
\end{tabular}
\end{center}
\vspace{10pt}

For a spacetime with no horizon, we choose the integration region of \eqref{eq:NelsonThm2} to extend from the origin to infinity. The boundary at infinity gives zero by the assumption of asymptotic flatness.
For the $(0,0)$ and $(1,-1)$ families, as the inner integration boundary is taken towards the origin, $r \to 0$, the inner boundary terms also tend to zero, and we consequently learn that if there is no source between the origin and infinity then one must have $R=0$ throughout spacetime. The $(0,0)$ family contains an $R=0$ solution, as does the $(1,-1)$ family. 
For the $(2,2)$ family, the boundary term blows up as $r \to 0$ and one can make no conclusion about the necessary vanishing of $R$. Note, however, that the $(2,2)$ family also does contain an $R=0$ solution, obtained by applying two constraints to its six free parameters. Compare this to the expression for the Ricci scalar in the linearised theory \eqref{eq:linricci} and to the analysis of the full theory given in Section \ref{sec:betazero}, where we found that the $\beta=0$ theory has four free parameters, in order to see that the $R=0$ condition imposes two parametric constraints on generic static spherically symmetric solutions.

The $R=0$ solution for the $(0,0)$ family is obtained if and only if $b_2=a_2$:
\begin{align}
A(r) & = 
1
+a_2 \left(
r^2
+r^4 \frac{12 \alpha  a_2+\gamma }{20 \alpha }
+r^6 \frac{320 \alpha ^2 a_2^2+100 \alpha  a_2 \gamma +\gamma ^2 }{1120 \alpha ^2}
 + O(r^8)
 \right)
\\
\frac{B(r)}{b_0} & = 
1
+a_2  \left(
r^2
+r^4 \frac{24 \alpha  a_2+\gamma }{40 \alpha }
+r^6  \frac{960 \alpha ^2 a_2^2+144 \alpha  a_2 \gamma +\gamma ^2 }{3360 \alpha ^2}
 + O(r^8)
 \right)
\end{align}
and $R_{\mu\nu}=0$ is obtained if and only if $a_2=0$.
Since $A(r \to 0) \to 1$ for all of $(0,0)$ this solution must have an even number of horizons.

The $R=0$ solution for the $(1,-1)$ family is obtained if and only if $a_4 = \frac{5}{3} a_1 b_2 - a_1^4$:
\begin{align}
A(r) & = 
a_1 r
-a_1^2 r^2
+a_1^3 r^3
+r^4 \left(\frac{5}{3}a_1 b_2-a_1^4\right)
+r^5 \left(a_1^5-\frac{23}{6} a_1^2 b_2\right)
+O(r^6) \\
\frac{B(r)}{b_{-1}} & = 
\frac{1}{r}
+a_1
+b_2 r^2
+\frac{1}{6} a_1 b_2 r^3
-\frac{1}{5} r^4 a_1^2 b_2
+O(r^5)\,.
\end{align}
The free parameter $a_1$ corresponds to the Schwarzschild mass and the free parameter $b_2$ controls the deviation from the Schwarzschild solution. Specifically, one has pure Schwarzschild, \ie $R_{\mu\nu}=0$, if and only if $b_2=0$, and inspection of the solution then shows that $a_1 = -\frac12(M_{\text{Schwarzschild}})^{-1}$, so we expect the solution to have a horizon for $a_1 < 0$ and to have no horizon for $a_1 > 0$.

The $(2,2)$ family (which does not appear in general relativity) has a four-parameter family of $R=0$ solutions but
cannot have $R_{\mu\nu} = 0$ for all $r$, because, \egns, in the $(2,2)$ family one has $R_{rr} = 3 r^{-2} + O(r^{-1})$.

A summary of boundary structure and parameter counts for the three near-origin solution families is given in Table \ref{tab:boundary&params}.

%\subsubsection{Summary}\label{subsubsec:summary}

\begin{table}[H]
\centering
\caption{Trace no-hair boundary structure and solution parameter counts\\ (including the trivial time-rescaling parameter)\label{tab:boundary&params}}
%\begin{tabular}{|c|c|c|c|}
%\hline
%$(s,t)$ solution family & $R\partial_rR$ as $r \to 0$ & free params. & free params. when $R=0$ \\
%\hline
%$(0,\;\; 0)$		&	vanishes		&	3	&	2\\
%$(1,-1)$			&	vanishes		&	4	&	3\\
%$(2,\;\; 2)$		&	blows up		&	6	&	4\\ %$O(r^{-1})$
%\hline
%soln.\  with horizon			&	finite at horizon				&	4	&	3\\
%\hline
%\end{tabular}
%\end{table}
\begin{tabular}{|c|c|cc|}
\hline
\rule{0pt}{2.6ex}
$(s,t)$ solution family
 & $\sqrt{AB} \, r^2 R \, \partial^r R$
 & \multicolumn{2}{c|}{number of free parameters} \\[1ex]
\cline{3-4}
\rule{0pt}{2.6ex}
% && \multicolumn{1}{c|}{(generic $\alpha,\beta$)} & \phantom{aa}($\beta=0$)\phantom{aa} \\[1ex]
 && \multicolumn{1}{c|}{generic theory} & $\beta=0$ theory \\[1ex]
\hline
\rule{0pt}{2.6ex}
$(0, \phantom{-} 0)$ &$O(r^3)   $& \multicolumn{1}{c|}{3} & 2 \\[1ex]
\rule{0pt}{2.6ex}
$(1,          -  1)$ &$O(r^3)   $& \multicolumn{1}{c|}{4} & 3 \\[1ex] 
\rule{0pt}{2.6ex}
$(2, \phantom{-} 2)$ &$O(r^{-1})$& \multicolumn{1}{c|}{6} & 4 \\[1ex]
\hline 
\end{tabular} 
\end{table}

%\section{Expansion around a horizon}\label{sec:horizonexp}
\section{Expansion around a nonzero radius $r_0$}\label{sec:horizonexp}
We can refine our understanding of the various forms of solutions by expanding now around an arbitrary radius $r_0$.
This will be easier to do in a slightly different set of variables:
\begin{equation}\label{eq:metricUsingf}
ds^2 = - B(r)\,  dt^2 + \frac{dr^2}{f(r)} + r^2 d\Omega_2^2 \,,
\end{equation} related to the usual Schwarzschild variables \eqref{eq:metric} by $A(r)=1/f(r)$.

We can use a Frobenius ansatz for the expansion about $r_0$ similar to our expansion \eqref{Frobseries} about $r=0$:
\bse\label{eq:aroundr0FrobAnsatz}
\begin{align}
f = & 
f_w (r-r_0)^w + f_{w+1} (r-r_0)^{w+1} + \dots\label{eq:aroundr0Frobf}
\\
\frac{B}{b_t} = & 
(r-r_0)^t + b_{t+1} (r-r_0)^{t+1} + \dots\label{eq:aroundr0FrobB}
\end{align}
\ese
for some exponents $w$ and $t$, not confusing these undetermined $(w,t)$ exponents with the undetermined $(s,t)$ exponents used earlier in the expansion \eqref{Frobseries} around $r=0$.
We shall find various Frobenius solution families and also some non-Frobenius families of solutions. However, detailed discussion of all of these is beyond the scope of this paper. 
In Section \ref{sec:aroundr0ExpansionsSummary} we shall present a summary and discussion of all the solutions that we have found.
For the purposes of the main thread of our discussion the situation where the space-time has a horizon will be of most interest to us and we turn to it now. 

\subsection{Solution family around a horizon}
%\subsection{Expansion around a horizon in detail}
We now focus on the properties of spherically-symmetric solutions with horizons. 
We saw in Section \ref{sec:tracenohair} that in a static asymptotically flat spacetime the presence of a horizon implies that the Ricci scalar must vanish and consequently the system becomes equivalent to Einstein-Weyl gravity, with Lagrangian
%%%%%
\begin{equation}
e^{-1}{\cal L} = 
\gamma \,  R - 
\alpha C^{\mu\nu\rho\sigma} C_{\mu\nu\rho\sigma}\,.
\end{equation}
%%%%%

%In order to study solutions near a horizon, we write the metric in the form
%\begin{equation}
%ds^2 = - B(r)\,  dt^2 + \fft{dr^2}{f(r)} + r^2 d\Omega_2^2\,,
%\end{equation}
%%%%%%
%\ie we let $A(r)=1/f(r)$.
Using the metric \eqref{eq:metricUsingf}
the equations of motion then imply
\bse\label{afeqns}
\begin{align}
0 = & 
r^2 \left( B \left(2 f B''+B' f'\right)-f B'^2\right)
\notag\\ &
+4   B \left(r f B'+B \left(r f'+f-1\right)\right)
\label{eq:afeqn1}
\\
0 = & 
\alpha  \Bigg(
r^3 f B' \left[ -B B' f'+f B'^2-2 B^2 f''\right]
\notag\\&
-2 B^2 f \left(r^2 B' f'-2 B \left(r^2 f''-r f'+2\right)\right)-B f^2 \left(3 r^2 B'^2+8 B^2\right)
\notag\\&
-r B^3 f' \left(3 r f'-4\right)
\Bigg)
%\notag\\&
+2 \gamma  r^2 B^2 \left(r f B'+B (f-1)\right)\,.
\label{eq:afeqn2}
\end{align}
\ese

%%%%%
   The Schwarzschild metric itself is of course still a solution to
Equations \eqref{afeqns}, with
%%%%%
\be
B(r)=f(r) = 1 -\fft{r_0}{r}\,. 
\ee
%%%%%

In the higher-derivative theory we do not have a general solution in closed form but a Frobenius analysis performed around the horizon can reveal its relation to the Schwarzschild solution. We consider solutions in the neighbourhood of a horizon, assumed to be at $r=r_0$.
%By definition, the metric function $B(r)$ vanishes at a horizon so we assume, 
By definition, the metric function $B(r)$ vanishes at a horizon so we look for solutions of the form\footnote{Note that an expansion of the form \eqref{eq:sqrtexpansions} for an asymptotically flat solution with a horizon, \ie a $\left(\frac{3}{2},\half\right)_{\sqrt{r-r_0}}$ expansion in the notation of Table \ref{tab:gen_r_0expfamilies} is not possible, as one must have vanishing Ricci scalar for asymptotically flat solutions and such an expansion does not then exist.} \eqref{eq:aroundr0FrobAnsatz} with $t>0$.
%\footnote{As in our analysis of the asymptotic behaviour of solutions near the origin given in Section \ref{sec:solutionsNearTheOrigin}, one might again worry that, when faced with such a nonlinear system of equations, standard Frobenius analysis about a horizon might not capture all possible solution behaviours. In the absence of any general theorems governing this case, we have explored a variety of other expansion possibilities, \eg involving logarithms or expansion in fractional powers of $r$. These have not revealed any expansion behaviours other than those of (\ref{eq:horFrobB}, \ref{eq:horFrobf}).}
%similarly to \eqref{Frobseries} for expansion about the origin, that
%\begin{equation}
%\frac{B}{b_t} = (r-r_0)^t + b_{t+1} (r-r_0)^{t+1} + \cdots\label{eq:horFrobB}
%\end{equation}
%where $t > 0$, and then again similarly for $f(r)$
%\begin{equation}
%f = f_s (r-r_0)^s + f_{s+1} (r-r_0)^{s+1} + \cdots\label{eq:horFrobf}
%\end{equation}
%for some $s$.
Using the expansions of $B$ and $f$ as given in \eqref{eq:afeqn2} shows that $w \leq \frac{3}{2}$. In \eqref{eq:afeqn1}, for $w < \frac{3}{2}$ it is the terms in the first line that contribute the leading-order term in $(r-r_0)$, and the reader can easily check that these vanish only if $t=2-w$. Using $t=2-w$ in \eqref{eq:afeqn2} then shows that the first line contributes the leading order term in $(r-r_0)$ and this then vanishes only if $w=t=1$. Thus both $B$ and $f$ must be linear in $(r-r_0)$ at the horizon, as in the Schwarzschild case. %Consequently, asymptotically flat solutions with a horizon must fall into the $(w,t)=(1,1)$ family of Table \ref{tab:gen_r_0expfamilies}.

The solution to the equations of motion then is found to have three free parameters: $b_1,f_1,r_0$. The other coefficients $b_n, f_m$ can be solved for in terms of these:
\bse
\begin{align}
f = & 
f_1 (r-r_0) 
+\left(  \frac{3 \gamma }{8 \alpha }-\frac{3 \gamma }{8 \alpha  f_1 r_0}+\frac{-2 f_1}{r_0}+\frac{1}{r_0^2}  \right) (r-r_0)^2 
+ O\left( (r-r_0)^3 \right)\,,
\\
\frac{B}{b_1} = & 
(r-r_0)
+\left( -\frac{\gamma }{8 \alpha  f_1}+\frac{\gamma }{8 \alpha  f_1^2 r_0}+\frac{1}{f_1 r_0^2}-\frac{2}{r_0} \right) (r-r_0)^2 
+ O\left( (r-r_0)^3 \right)\,,
\end{align}
\ese
and so on. 
The parameter $b_1$ is trivial, as we have seen, in the sense that it can be absorbed into 
a rescaling of the time coordinate.  
Note that the Schwarzschild solution corresponds to the case where
$f_1=1/r_0$, with the same time coordinate as for Schwarzschild if one also chooses $b_1=1/r_0$.

This count of the free parameters tells us which of the three families near the origin corresponds to asymptotically flat solutions with horizons. Consider the $(s,t)=(1,-1)$ family of Table \ref{tab:boundary&params}. As we have seen in Section \ref{sec:tracenohair}, for asymptotically-flat solutions this family must have $R=0$ and its equations of motion accordingly must become equivalent to the $\beta = 0$ theory as considered in this section.  In the $\beta = 0$ theory, one accordingly has three free parameters as shown in Table \ref{tab:boundary&params}. The theory thus contains the two-parameter Schwarzschild solution, together with a one-parameter family of deviations from Schwarzschild.  Nearby the Schwarzschild solution within this one-parameter family of deviated solutions, we would certainly expect the horizon still to be present. So we expect the $(1,-1)$ family to be a three-parameter family, in which a three-dimensional volume of that parameter space has a horizon. Therefore in the $\beta = 0$ theory we identify the $(s,t)=(1,-1)$ indicial solution family obtained from expansion near the origin with the solution family containing horizons.
 
One needs to be careful with the logic here. Although we say that asymptotic flatness implies $R=0$ for the $(s,t)=(1,-1)$ solution family, and therefore its equivalence to solutions of the Einstein-Weyl theory, the converse is not necessarily true. 
It is likely that not all $(1,-1)$ solutions of Einstein-Weyl gravity are asymptotically flat, and in fact the loss of asymptotic flatness is a natural guess for the consequence of turning on the `non-Schwarzschild' parameter measuring the deviation from Schwarzschild. 
In Section \ref{sec:Schlinnohair}, we will further explore such deviations from general relativity in Einstein-Weyl gravity and the implications for their asymptotic behaviour through a linearised expansion in the non-Schwarzschild parameter.

\subsection{Summary of expansion behaviours around a non-zero radius $r_0$}\label{sec:aroundr0ExpansionsSummary}
The Frobenius ansatz \eqref{eq:aroundr0FrobAnsatz} has three solution families. 
The first is simply the $(0,0)$ family, corresponding to no special radius. Its main value is reinforcing our conclusions about the generic number of free parameters of the theory. 
The second is the $(1,1)$ family already discussed, describing a horizon.
The third is the $(1,0)$ family, which we shall describe as a wormhole in Section \ref{sec:numanalysis}.

As mentioned in our Section \ref{sec:solutionsNearTheOrigin} discussion of expansions around the origin, a natural concern is that there might be other solution behaviours not captured by the integral-step Frobenius expansion. We cannot check all alternative expansions exhaustively, but we have tried a variety of non-Frobenius expansions and have found two other solution families. 
Both of these non-Frobenius families involve half-integer as well as integer powers of $(r-r_0)$. As before, we denote Frobenius families by bracketed pairs of indices $(w,t)$, and we shall denote the new series similarly but with a subscript: $(w,t)_{\sqrt{r-r_0}}$, where they go as
\bse\label{eq:sqrtexpansions}
\begin{align}
f = & 
f_0 (r-r_0)^w 
+ f_1 (r-r_0)^{w+\half} 
+ f_2 (r-r_0)^{w+1}
+ \dots
\\
\frac{B}{b_0} = & 
(r-r_0)^t 
+ b_1 (r-r_0)^{t+\half} 
+ b_2 (r-r_0)^{t+1} 
+ \dots
\end{align}
\ese

We shall not give a detailed analysis of all five solution families in this paper, but we summarise our findings in in Table \ref{tab:gen_r_0expfamilies}.

\begin{table}[H]
\centering
\caption{General expansion behaviours around a nonzero radius $r_0$
\label{tab:gen_r_0expfamilies}}
\begin{tabular}{|c|c|cc|}
\hline
\rule{0pt}{2.6ex}
$(w,t)$ solution family
 & $\sqrt{AB} \, r^2 R \, \partial^r R$
 & \multicolumn{2}{c|}{number of free parameters} \\[1ex]
\cline{3-4}
\rule{0pt}{2.6ex}
 && \multicolumn{1}{c|}{generic theory} & $\beta=0$ theory \\[1ex]
\hline
\rule{0pt}{2.6ex}
$(0, 0)$						& $O( 1 )         $ & \multicolumn{1}{c|}{6} & 4 \\[1ex]
\rule{0pt}{2.6ex}
$(1, 1)$						& $O(r-r_0)$ & \multicolumn{1}{c|}{4} & 3 \\[1ex] 
\rule{0pt}{2.6ex}
$(1, 0)$						& $O(\sqrt{r-r_0})$ & \multicolumn{1}{c|}{3} & 2 \\[1ex]
\rule{0pt}{2.6ex}
$(1, 0)_{\sqrt{r-r_0}}$			& $O( 1 )          $ & \multicolumn{1}{c|}{6} & 4 \\[1ex]
\rule{0pt}{2.6ex}
$\left(\frac{3}{2},\half\right)_{\sqrt{r-r_0}}$ & $O(\sqrt{r-r_0})$ & \multicolumn{1}{c|}{3} & N/A \\[1ex]
\hline 
\end{tabular} 
\end{table}
Note that the $\left(\frac{3}{2},\half\right)_{\sqrt{r-r_0}}$ solution family does not occur in the $\beta=0$ theory. The $(1, 0)$ solution family is a subset of the $(1, 0)_{\sqrt{r-r_0}}$ family, obtained by setting the coefficient of odd powers of $\sqrt{r-r_0}$ to zero while holding the coefficients of even powers at finite values.

\section{\protect\scalebox{.93}[1]{No-hair theorem for a linearised deviation from Schwarzschild}}\label{sec:Schlinnohair}

   We saw in Section \ref{sec:tracenohair} that by considering the trace of the
field equations for gravity with general quadratic curvature terms added, one 
can derive a no-hair theorem that shows that the Ricci scalar must vanish
in any asymptotically-flat black hole solution.  Unfortunately, 
similar arguments applied to the full set of field equations fail to
establish a more powerful result that one might have hoped to demonstrate, namely
the vanishing of the full Ricci tensor for all asymptotically-flat spherically-symmetric
solutions with horizons.  Had one been able to obtain such a result, this would have
shown the Schwarzschild solution to be the unique static 
spherically-symmetric asymptotically-flat black-hole solution in theories of gravity
with curvature-squared corrections.

Indeed, as we found in Ref.\ \cite{Lu:2015cqa}, there {\em are} non-Schwarzschild black-hole solutions to be found numerically, so the failure of a full Lichnerowicz-type no-hair theorem including the traceless components of the field equations \eqref{generalEOM} is now seen to have been quite indicative. Nonetheless, one may still obtain useful information about the set of solutions with horizons from a no-hair theorem analysis carried out to linearised order in the 
non-Schwarzschild parameter discussed in the last section.  
As we shall see, the upshot from this analysis is that, provided the curvature-squared terms have sufficiently small coefficients in comparison
to the scale size of the black hole, then there can be no well-behaved
static and spherically-symmetric black holes that are perturbatively 
close to the
Schwarzschild solution. So in this restricted sense, one can show that 
the Schwarzschild solution is generally an {\em isolated} solution, 
discretely separated from other asymptotically-flat solutions with horizons.
 
   We may consider solutions of the equations (\ref{afeqns}) that
are infinitesimally close to Schwarzschild by writing
%%%%%
\be
B(r) = 1- \fft{r_0}{r\left(1 +\epsilon Z_B(r)\right)}\, \qquad
\frac{1}{A(r)} = f(r)= 1 -\fft{r_0}{r(1+ \epsilon Z_A(r))}\,,\label{afpara}
\ee
%%%%%
and keeping only terms of order $\epsilon$.  
From the two coupled equations of motion in $Z_A$ and $Z_B$, a single 3rd-order ordinary differential equation purely for $Z_A(r)$ can be obtained by eliminating $Z_B(r)$:
\bea
Z_B(r) - (r-r_0)Z_B'(r) &=& 
Z_A(r)+\frac{\alpha  \left(-8 r^2+16 r r_0-9 r_0^2\right) (r-r_0) }{2 \gamma  r^4-2 \gamma  r^3 r_0-4 \alpha  r r_0+5 \alpha  r_0^2}Z_A'(r)\cr
&&\null\hspace{1cm} +\frac{2 \alpha  r (2 r-3 r_0) (r-r_0)^2 }{2 \gamma  r^4-2 \gamma  r^3 r_0-4 \alpha  r r_0+5 \alpha  r_0^2}Z_A''(r)\,.
\eea
In fact, the resulting equation in $Z_A$ involves only $Z_A'$, $Z_A''$ and $Z_A'''$
terms, and consequently we have a 2nd-order ordinary differential equation for $Z_A'$.  It is useful to 
introduce a new variable $Y(r)$, defined by
%%%%%
\be
Z_A(r) = \int_{r_0}^r Y(\tilde r)\omega(\tilde{r}) d\tilde r\,,\label{ZtoY}
\ee
%%%%%
where $\omega(r)=1$ for now but this will be revised later. The lower limit in (\ref{ZtoY}) is chosen to be $r_0$ in order to ensure that $Z_A(r)$ vanishes on the horizon.\footnote{The
third linearly independent solution of the 3rd-order ordinary differential equation for $Z_A(r)$
itself is $Z_A(r)=\,\hbox{const.}$, which, as can be seen from
(\ref{afpara}), just describes a perturbation that shifts the
mass of the Schwarzschild solution by an infinitesimal constant.}  
Using the abbreviation 
\be
\zeta = \alpha (\gamma r_0^2)^{-1}\,,\label{zeta}
\ee
the second-order ordinary differential equation
for $Y(r)$ is then 
%%%%%
\be
h_0 \, Y + h_1\, Y' + h_2\, Y''=0\,,\label{Yeqn}
\ee
%%%%%
where one has
%%%%%
\begin{align}
h_0 = &\, 
2 r^7-2 r_0 r^6-8 r_0^2 r^5 \zeta+16 r_0^3 r^4 \zeta-5 r_0^4 r^3 \zeta-32 r_0^5 r^2 \zeta^2+44 r_0^6 r \zeta^2-20 r_0^7 \zeta^2
\,,\cr
h_1 = &\,
4 r^2 \left(2 r-3 r_0\right) r_0^2 \zeta \left(r^3-r_0 r^2+r_0^3 \zeta\right)
\,,\cr
h_2 = &\, 
-2 r^2 \left(r-r_0\right) r_0^2 \zeta \left(2 \left(r-r_0\right) r^3+r_0^3 \left(5 r_0-4 r\right) \zeta\right)
\,.
\label{hexp}
\end{align}
%%%%%

   One can easily see from (\ref{Yeqn}) that at large $r$ the two solutions to the field equations 
go like
\begin{equation}
Y \sim 
  a_1 \left( 1+m_2 r\right) e^{- m_2 r}
+ a_2 \left( 1-m_2 r\right) e^{  m_2 r}
\end{equation}
where $m_2=\sqrt{\gamma/2\alpha}$ as before.  
For $r$ close to the horizon at $r=r_0$,
one can once again use the Frobenius method to find the $r\to r_0$ asymptotic behaviour of 
the two independent solutions $Y_1$ and $Y_2$.  We find that they 
take the asymptotic forms
%%%%%
\bea
Y_1 &=& 1 + c_1\, (r-r_0) + c_2\,(r-r_0)^2 + c_3\, (r-r_0)^3 \cdots\,, 
\qquad c_i=c_i(\alpha,r_0)\ 
\hbox{for}\ i\ge1\,,\cr
Y_2 &=& Y_1\, \log(r-r_0) + \fft{b_{-1}}{r-r_0} + b_1\, (r-r_0) + 
b_2\, (r-r_0)^2 + b_3\, (r-r_0)^3 + \cdots\,,\cr
&& \hbox{with}\ b_{-1}=-\fft{\alpha}{r_0}\,,\qquad b_i=b_i(\alpha,r_0)\
\hbox{for}\ i\ge  1\,.
\eea
%%%%%
Thus, in order for the metric perturbation $Z_A(r)$ to be non-singular at large $r$
we must have $a_2=0$, while for non-singular behaviour near
$r=r_0$ the overall coefficient of the $Y_2$ solution
must be zero.  We shall now attempt
to show that no such solution $Y$ that interpolates between these limiting
forms can exist. 

   To do this, we take the $Y$ equation (\ref{Yeqn}),
multiply it by $u(r) Y(r)$ for some chosen $u(r)$ and then integrate the result from $r=r_0$ 
(the horizon) out to infinity.  Firstly, we note that
%%%%%
\be
0=(h_0 \, Y + h_1\, Y' + h_2\, Y'') u Y =
  u h_0 Y^2 - u h_2 {Y'}^2 + (u h_1 - u' h2 - u h_2')\, Y Y' 
     +( u h_2 Y Y')'\,.\label{poseqn}
\ee
%%%%%
We now choose $u(r)$ so that the coefficient of $Y Y'$ vanishes, by solving
%%%%%
\be
u h_1 - u' h_2 - u h_2' =0\,.
\ee
%%%%%
This gives, up to a constant factor that we may take to be 1, 
%%%%%
\be
u(r) = 
\frac{\left(r-r_0\right)}{\left(2 \left(r-r_0\right) r^4+r_0^3 \left(5 r_0-4 r\right) r \zeta\right)^2}\,.
\label{usol}
\ee
%%%%%
    Integrating (\ref{poseqn}) from $r_0$ out to infinity, we thus obtain
%%%%%
\be
\int_{r_0}^\infty dr \big( u h_0\, Y^2 - u h_2\, {Y'}^2\big)=0\,.
\ee
%%%%%
Note that we can drop the total derivative term, since we have established that an
acceptable solution for $Y$ must satisfy $Y\sim\,
\hbox{const.}$ near $r=r_0$ and $Y\sim e^{-mr}$ near infinity.  If we can
show that $u h_0$ and $-u h_2$ are both non-negative in the interval
$r_0\le r\le\infty$, then we will have shown that no acceptable 
solution $Y$ can exist.

  The function $u(r)$ obtained in (\ref{usol}) is manifestly non-negative 
in the range $r_0\le r\le\infty$.  
It is then evident from (\ref{hexp}) that showing the non-negativity of
$u h_0$ and $-u h_2$ is equivalent to showing that $h_0$ and $H_2$, 
given by
%%%%%
\bea
h_0 &= &
2 r^7-2 r_0 r^6-8 r_0^2 r^5 \zeta+16 r_0^3 r^4 \zeta-5 r_0^4 r^3 \zeta-32 r_0^5 r^2 \zeta^2+44 r_0^6 r \zeta^2-20 r_0^7 \zeta^2
\,,\cr
H_2 &= & 
\frac{h_2}{-2 r^2 \left(r-r_0\right) r_0^2 \zeta} = 
2 \left(r-r_0\right) r^3+r_0^3 \left(5 r_0-4 r\right) \zeta \label{P0P1}
\eea
%%%%%
are non-negative in the interval $r_0\le r\le \infty$, for some range of $\zeta \geq 0$.

   It is easy to see that $H_2$ is non-negative in the whole 
interval if and only if 
%%%%%
\be
0\le \zeta \le \fft{27}{8}\, .\label{plim}
\ee
%%%%%
The non-negativity of $h_0$ provides a stronger condition on $\zeta$. 
Setting $r=r_0$, we see that $h_0(r_0)= r_0^7 \left(3-8 \zeta\right) \zeta $, and so we must certainly have $0\le \zeta \le \frac{3}{8}$.  
An easy way to investigate the bound on $\zeta$ under which $h_0(r)$ is non-negative
in the entire range $r_0\le r\le\infty$ is to write
%%%%%
\be
r=r_0\, (1+x)\,,\qquad \zeta= \fft{ \zmax}{1+y}\,,
\ee
%%%%
since then the ranges $r_0\le r\le\infty $ and $0\le \zeta \le \zmax $ are mapped 
into the positive quadrant $0\le x\le \infty$, $0\le y\le \infty$. 
We then bring the expression for $h_0$ over a common denominator 
(which is the manifestly positive quantity $(1+y)^2$), and examine the
numerator, which is a multinomial in $x$ and $y$ \ie it is of the form
\begin{equation}
\sum_n^N \sum_m^M C_{n,m} \,\, x^n \, y^m   \label{eq:pertNoHrSuffCond}
\end{equation}
where the $C_{n,m}$ are functions of $r_0$ and $\zmax$. The condition $C_{n,m} \geq 0$ for all $n,m$ is clearly sufficient (but may not be necessary) for non-negativity of $h_0$ in $r_0\le r\le \infty$. This condition easily yields the bound
%%%%%
\be
 0\le \zeta \le \frac{3}{8} \,.\label{pbound}
\ee
%%%%%
Conversely, as we have seen, if $\zeta$ exceeds this bound then
$h_0$ will be negative at $r= r_0$.

   The upshot from this discussion is that, provided
$\zeta$ is bounded from above by (\ref{pbound}), then there cannot exist
a regular infinitesimal perturbation of the metric away from the
Schwarzschild solution.  For $\zeta$ exceeding this bound, we can gain
no information from this discussion. 

One can actually improve the upper bound on $\zeta$ by making a different choice of $\omega(r)$ in \eqref{ZtoY}, and by repeating the previous steps. 
For example, if we take $\omega(r)=(c r_0 + r)^{-1}$,
where $c > -1$ is a constant to be chosen, 
the optimal
bound $\zmax$ is the largest positive root of the sextic
%%%%%
\begin{equation}
9600 \, \zmax^6+8624  \, \zmax^5-9360  \, \zmax^4-4461  \, \zmax^3+1216  \, \zmax^2+1116  \, \zmax-48
\, ,
\end{equation}
%%%%%
which is approximately given by $\zmax \sim 0.617292 $.  
This is achieved by choosing the constant $c$ to be given by
%%%%%
\be
c= 
\frac{-36  \, \zmax^2+19  \, \zmax+2}{20  \, \zmax^2-9  \, \zmax-2}
\sim 0.164789 \,.
\ee
%%%%%

A slight improvement on this can be achieved by taking instead $\omega(r)=(c r_0^3+r^3)^{-1/3}$.  
We then find $0 \leq \zeta \leq \zmax$ with $\zmax$ the largest positive root of 
\begin{equation}
28160  \, \zmax^4+12176  \, \zmax^3-43374  \, \zmax^2+19179  \, \zmax-2322 = 0\,,
\end{equation}
which is $\zmax \sim 0.6262615$, attained when the constant %0.62626146871
$c$ is chosen to be given by 
\begin{equation}
c = \frac{3-4  \, \zmax}{8  \, \zmax-3} \sim 0.2462346 \,.%0.24623459369
\end{equation}

The condition of $C_{n,m} \geq 0$ in \eqref{eq:pertNoHrSuffCond} is a sufficient condition, and the necessary bound on $\zeta$ may be better. 
Trying different functions $\omega(r)$ could presumably improve the bound further. 
The best one could hope to achieve by this method is the bound (\ref{plim}) (valid for any function $\omega(r)$) arising from the need for $H_2$ to be non-negative also.  
In any case, we have established that, provided $\zeta$ is sufficiently small, there is a linearised no-hair theorem that rules out regular black holes that are close to the Schwarzschild solution.

The treatment of
the traceless components of the field equation by such 
a linearised perturbative approach establishes, for $\zeta$ appropriately 
bounded, that the Schwarzschild solution is {\em isolated}: there can be 
no other nearby asymptotically-flat solutions with horizons. The notion 
of `nearby' solutions is made clear by our general analysis of the 
parametric dependence of solutions with horizons. 
From the parameter count 
summarised in Section \ref{subsubsec:summary}, we presented an argument in Section \ref{sec:horizonexp} for identifying the 3-parameter family of solutions with horizons with the 3-parameter $(1,-1)$ family of solutions around the origin.
The classic 2-parameter Schwarzschild solution is clearly contained in
this family. 
It is in terms of the single remaining parameter of the $(1,-1)$ family 
that one can characterise the `distance' of other nearby solutions with 
horizons from the Schwarzschild solution.

The need to make such a cautious statement about the isolation of the 
Schwarzschild solution, however, clearly raises questions as to 
whether there might exist other asymptotically-flat solutions with 
horizons that are not `near' to Schwarzschild. From Ref.\ \cite{Lu:2015cqa}, 
we know this in fact to be the case. In the next section, 
we give a brief overview of what can be said about such solutions from 
numerical studies.

\section{Numerical analysis}\label{sec:numanalysis}

The detailed nonlinear field equations (given in Appendix \ref{app:fullequations}) for our Schwarzschild-coordinate spherically-symmetric system in the general $\alpha$, $\beta$, $\gamma$ theory \eqref{HDGaction} are clearly not very amenable to a closed-form solution. Having studied the asymptotic behaviour of solutions at the origin, at spatial infinity and at a horizon in Sections \ref{sec:solutionsNearTheOrigin}, \ref{sec:weakFieldExpansion} and  \ref{sec:horizonexp}, we now need to consider what happens in between these various limiting regions. This is only approachable by numerical study. We do not purport to give an exhaustive treatment of numerical solutions to the theory \eqref{HDGaction} here, but some review of what is already known and what can be obtained by Mathematica experimentation is in order \cite{alfio}.

\subsection{$(2,2)$ solutions}\label{ssec:22solsnum}

Firstly, let us consider solutions that could be obtained from coupling to a positive-energy shell source as discussed in Section \ref{sec:nonlinshellsources}. Only the $(2,2)$ indicial family has the full count of six parameters that are required to satisfy the six continuity and jump conditions across a shell source. After such matching, two parameters must implicitly be used to guarantee the absence of rising exponential behaviour at spatial infinity, corresponding to the rising spin-two and spin-zero Yukawa terms of the linearised theory. It is not known, however, which combinations of free parameters near the origin, given in Table \ref{tab:solparametercounts}, need to be tuned so as to eliminate the rising behaviour at spatial infinity. In order to match a $(2,2)$ family solution on to asymptotically-flat behaviour at infinity, one procedure is to start with a series-expanded solution near the origin and integrate outwards numerically, and also to start from an asymptotically-flat solution at spatial infinity and integrate inwards, then to adjust parameters so as to make the two numerical solutions match at an intermediate radius. Such a procedure was carried out in Ref.\ \cite{Holdom:2002xy} for the theory with $m_2=m_0$, which in the notation of this paper means $\alpha=3\beta$. Owing to the trace no-hair theorem as presented in Section \ref{sec:tracenohair}, any asymptotically-flat solution that has any amount of falling spin-zero ${e^{-m_0r}\over r}$ Yukawa behaviour near spatial infinity cannot have a horizon; as one can see from Eq.\ \eqref{eq:linearisedDeltaSourceRicci}, such solutions necessarily have $R\ne0$ in the $r\to\infty$ asymptotic region. Indeed, the asymptotically flat $(2,2)$ family solution found in Ref.\ \cite{Holdom:2002xy} displays a dominant $1/r$ Schwarzschild-type behaviour as $r\to\infty$, but it also displays a falling Yukawa correction and deviates strongly from Schwarzschild at smaller $r$. It does not encounter a horizon at intermediate $r$ values, but limits to $(2,2)$ family behaviour near the origin. The calculation of Ref.\  \cite{Holdom:2002xy} was made for a normal positive-sign mass $M$ (given \cite{ADMmass}  by $8\pi$ times the coefficient of $1/r$ in $g_{tt}$ as $r\to\infty$ for a theory with $\gamma=1$).

A similar calculation can be made in the $\gamma R -\alpha C^2$ theory with $\beta=0$, in which a vanishing Ricci scalar, $R=0$, is naturally guaranteed. Accordingly, this theory benefits from a further reduction of the set of third-order nonlinear field equations as given in Appendix \ref{app:fullequations} down to a pair of second-order equations. These are equivalent to the system \eqref{afeqns} for the $f=1/A$ and $B$ variables. This $\beta=0$ theory also has asymptotically flat, limiting to $(2,2)$ indicial, solutions without a horizon. A representative numerical solution with positive mass $M$ is shown in Figure \ref{fig:(2,2)toflat}.
\begin{figure}[h]
\centering
\includegraphics[width=210pt]{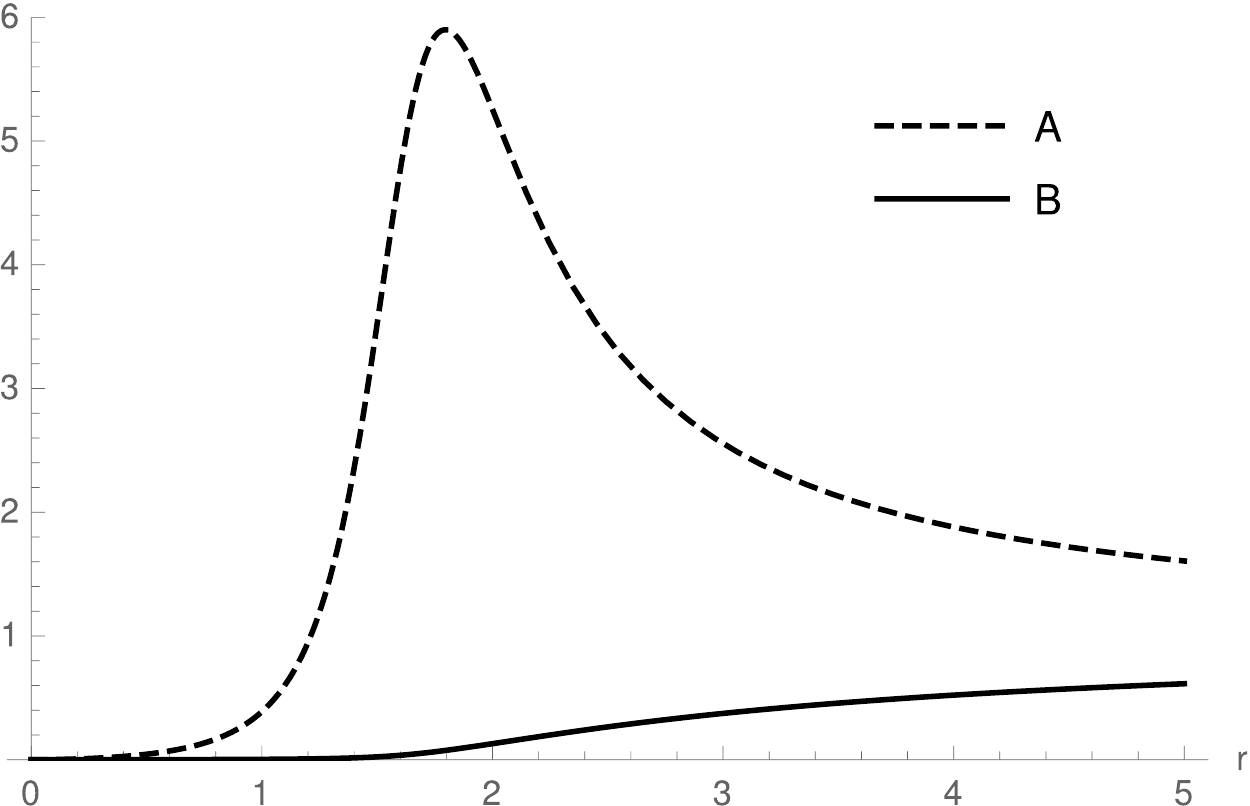}\hskip.4cm\includegraphics[width=210pt]{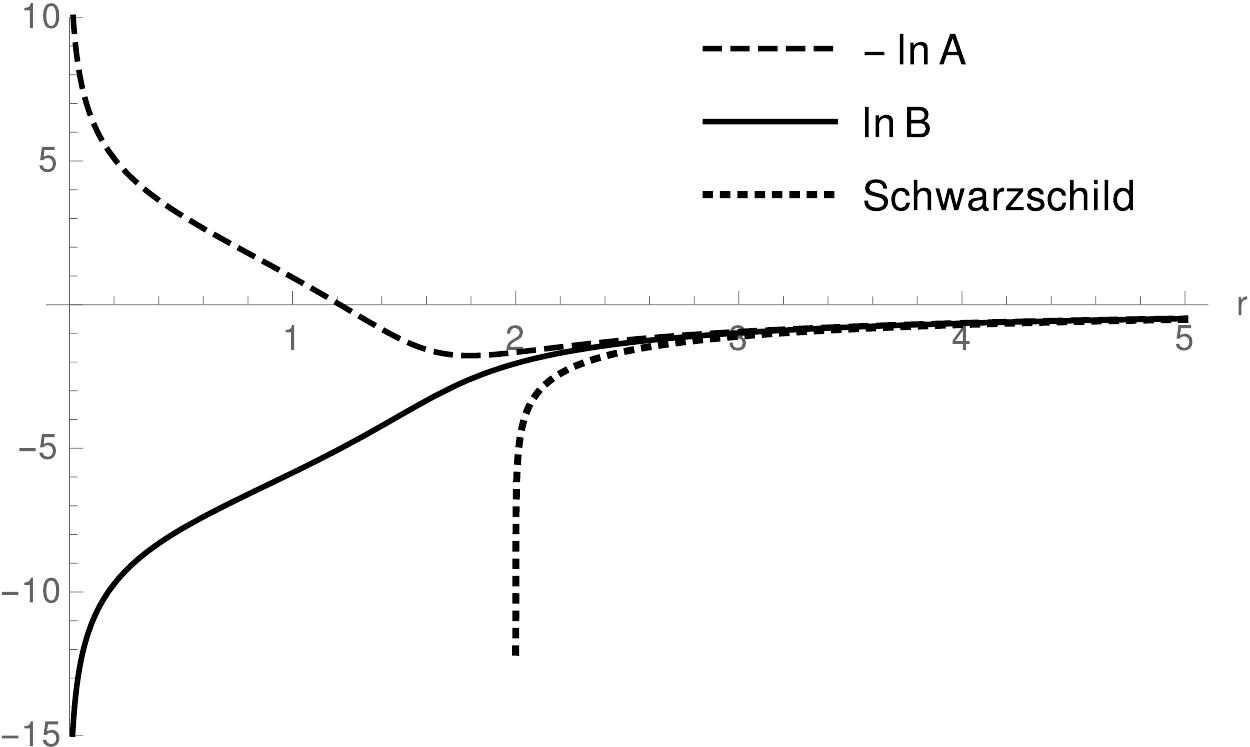}
\caption{{Horizonless asymptotically-flat solution at spatial infinity limiting to a $(2,2)$ indicial family solution at the origin. The right-hand plot for $-\ln A(r)$ and $\ln B(r)$ shows the relation to the Schwarzschild solution.}}
\label{fig:(2,2)toflat}
\end{figure}

The numerically obtained asymptotically flat $(2,2)$ solution shown in Figure \ref{fig:(2,2)toflat} was obtained for a positive ADM mass $M$ but with a negative $C_{2-}$ falling asymptotic Yukawa coefficient. Comparing to linearised theory solutions obtained with various sorts of distributional matter coupling as discussed in Section \ref{sec:weakFieldExpansion}, one sees that this relative sign between $M$ and $C_{2-}$ differs from that obtained for a point source in \eqref{eqa:linearisedDeltaSourceSol}. Recall, however, that for the higher-derivative theory \eqref{HDGaction} there is no Birkhoff theorem, and one does not have any expectation of a universal sign relation between $M$ and $C_{2-}$. Comparing instead to the linearised theory with the `balloon' source \eqref{eqa:balloonsource}, one finds that $C_{2-}$ can have either sign compared to $M$, so the sign found for $C_{2-}$ in the Figure \ref{fig:(2,2)toflat} solution poses no particular puzzle.

While in Ref.\ \cite{Holdom:2002xy} the method was to combine a shooting-out calculation from the origin with a shooting-in calculation from large radius, in order to find solutions like that of  Figure \ref{fig:(2,2)toflat} we employed a simpler method of just integrating inwards from a large radius using initial conditions taken from an asymptotically flat solution of the linearised theory. In general, this leads to divergent behaviour for $A(r)$ or $B(r)$  at small radii, but one should take care to notice that there are two kinds of divergent behaviour that can occur. Holding the linearised-theory's coefficient $C_{2,0}\sim -M$ fixed in \eqref{eqa:linearisedgensol} while varying the falling Yukawa coefficient $C_{2-}$, one finds ranges of $C_{2-}$ values for which $A(r)\to\infty$ while $B(r)\to0$ as $r\to0$, but then one finds a different range of $C_{2-}$ values for which $A(r)\to0$ while $B(r)\to\infty$. In between these ranges one finds (subject to numerical accuracy) a value of $C_{2-}$ for which both $A(r)\to0$ and $B(r)\to0$: this is a $(2,2)$ solution as shown in Figure \ref{fig:(2,2)toflat}.

This procedure for finding asymptotically flat $(2,2)$ solutions such as that of Figure \ref{fig:(2,2)toflat} reveals another feature of the overall solution space that calls for further study. If the $(2,2)$ indicial family solution lies on a separatrix in parameter space between two other kinds of more generic solution, what are these other kinds of solution? One of these other kinds of solution may be viewed as a `wormhole', to which we turn our attention next.

\subsection{Wormholes}\label{ssec:wormholesnum}

Another type of solution that can be found numerically may be described as a `wormhole'.\footnote{Wormholes have also recently been considered in the pure $R^2$ theory in Ref.\ \cite{Duplessis:2015xva}.} Such solutions are characterised by the existence of a zero for $f(r)=1/A(r)$ but with $B(r)=-g_{tt}\ne 0$. We have seen in Section \ref{sec:horizonexp} that solutions with $B(r)$ vanishing at some radius $r_0$ must also have $f(r)$ vanishing at $r_0$ as well. However, the converse is not necessarily true. Integrating inwards from an asymptotically flat solution at spatial infinity, one finds such solutions starting from a linearised solution \eqref{eqa:linearisedgensol} with chosen values of $C_{2,0}$ and $C_{2\,-}$. In this way, one finds solutions with either sign of $M=-8 \pi C_{2,0}$ and either sign of the large $r$ spin-two Yukawa coefficient $C_{2\,-}$.

Another way to investigate such solutions with $f(r_0)=0$ but $B(r_0)\ne 0$ is again to use Frobenius asymptotic analysis to find the possible behaviour as $r\to r_0$ and then to integrate outwards, looking for asymptotically flat solutions. Asymptotic analysis indeed shows, as one can see from the existence of the $(w,t)=(1, 0)$ and $(w,t)=(1, 0)_{\sqrt{r-r_0}}$ solution families shown in Table \ref{tab:gen_r_0expfamilies}, that there can be solutions for which $f(r)$ vanishes at some radius $r_0$ but where $B(r)$ remains at some nonzero value. We have studied this in particular in the $\gamma R -\alpha C^2$ theory with $\beta=0$.

As we have seen above in Section \ref{sec:aroundr0ExpansionsSummary} and summarised in Table \ref{tab:gen_r_0expfamilies}, asymptotic analysis as $r\to r_0$ for $f(r_0)=0$ but with $B(r_0)\ne 0$ turns up the following situation.
%\footnote{As in our analysis of the asymptotic behaviour of solutions near the origin given in Section \ref{sec:solutionsNearTheOrigin} and again in our analysis of solutions near a horizon given in Section \ref{sec:horizonexp}, one might worry that Frobenius power-series analysis might not capture all possible solution behaviours. Exploration of other non-standard expansion possibilities, \eg involving logarithms, has revealed no further consistent expansion systems for the $f(r_0)=0\,,\ B(r_0)\ne 0$ case, however.}
 The leading term in $f(r)$ is always linear in $(r-r_0)$ and the leading term in $B(r)$ is always, by assumption, a constant. Since we are also by assumption considering the first zero of $f(r)$ as $r$ comes in from infinity, without $B(r)$ having yet crossed zero (which would have constituted a horizon as we saw in Section \ref{sec:horizonexp}), the $B(r_0)=b_0$ constant must be positive. It is at this point that the half-integral-step expansions of type $(w,t)=(1, 0)_{\sqrt{r-r_0}}$ from Table  \ref{tab:gen_r_0expfamilies} become relevant. 
If half-integral steps are allowed, then one finds an expansion with four free parameters, which is the generic number of free parameters for spherically symmetric solutions  in the $\gamma R -\alpha C^2$ theory, which has two second-order field equations (equivalent to those given in \eqref{afeqns}) for $f(r)$ and $B(r)$.\footnote{A confirmation of this analysis may be observed in numerical solutions shooting inwards from asymptotically flat spacetime. For $f(r_0)=0$ but $B(r_0)>0$, there is an apparent divergence in the gradient of $B$ as one approaches $r_0$, where $f$ has a zero, agreeing with an expansion structure $f(r)=f_1(r-r_0)+f_{3/2}(r-r_0)^{3/2}+\ldots$ and $B(r)=b_0+b_{1/2}(r-r_0)^{1/2}+\ldots$.}

% 
% There now arises a new situation, which we have not found in the other Frobenius asymptotic analyses in this paper. The steps in powers of $(r-r_0)$ after the leading terms can either be integral or half-integral, \ie the solution family can be either of $(w,t)=(1, 0)$ or of $(w,t)=(1, 0)_{\sqrt{r-r_0}}$ expansion types. If half-integral steps are allowed (so that $f(r)=f_1(r-r_0)+f_{3/2}(r-r_0)^{3/2}+\ldots$ and $B(r)=b_0+b_{1/2}(r-r_0)^{1/2}+\ldots$), then, as shown in shown in Table \ref{tab:gen_r_0expfamilies}, one finds an expansion with four free parameters, which is the generic number of free parameters for spherically symmetric solutions  in the $\gamma R -\alpha C^2$ theory, which has two second-order field equations (equivalent to those given in \eqref{afeqns}) for $f(r)$ and $B(r)$.

\begin{figure}[H]
\centering
\includegraphics[width=200pt]{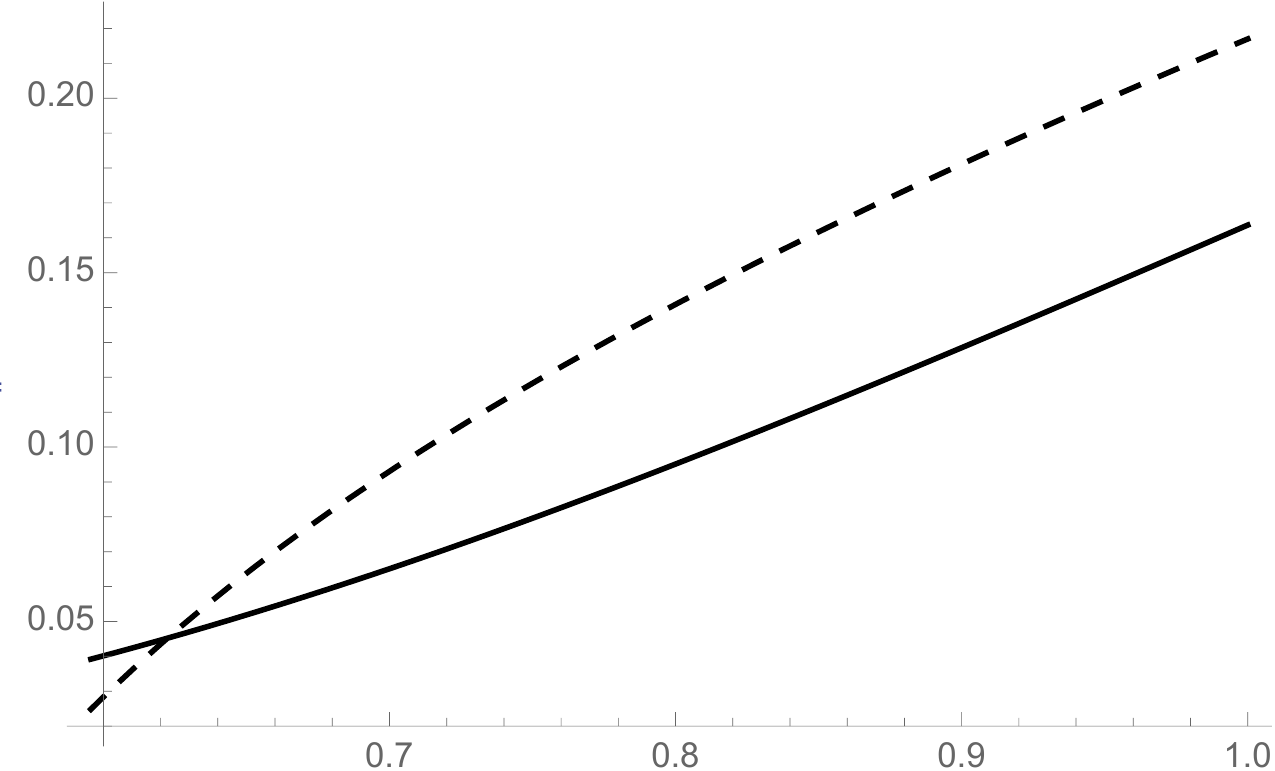}\ \ 
\includegraphics[width=200pt]{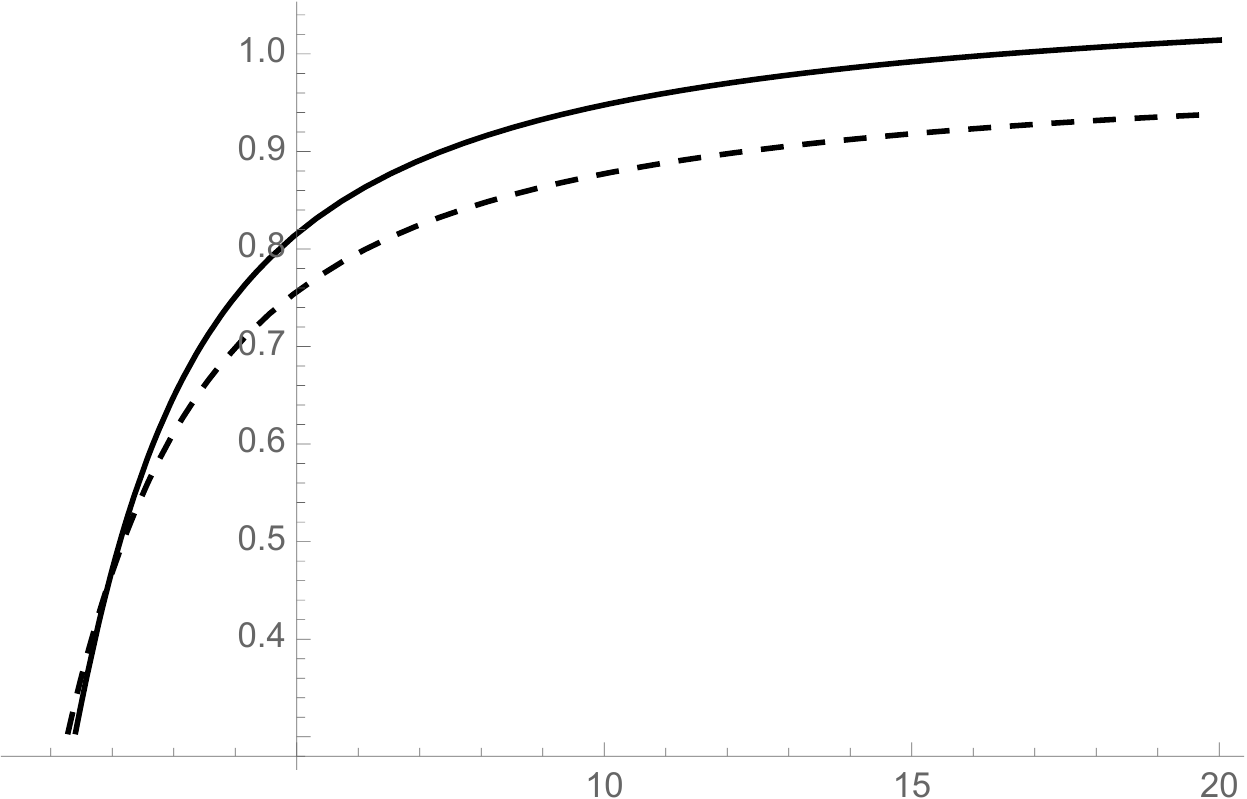}
\caption{$\Z_2$ symmetric wormhole solution with $f(r)\to 0$ as $r\to r_0$ (dashed line) and $B(r)\to\text{const.}$ (solid line). The left plot shows the small-scale structure near $r_0\simeq 0.57$ and the right plot shows larger-scale structure.}
\label{fig:wormholes}
\end{figure}

Assuming instead integral steps in powers of $(r-r_0)$ after an initial zero at $r_0$ leads to a more constrained solution system with only two free parameters: the trivial rescaling parameter affecting $B=-g_{tt}$ (which may be realised as $b_0$) and $r_0$ itself. Numerically integrating outwards in $r$, one then generally finds rapidly divergent behaviour as $r\to\infty$, but this behaviour can be of two different types, similarly to the two types of divergent behaviour surrounding the asymptotically flat $(2,2)$ solutions discussed in Section \ref{ssec:22solsnum}: one type has $f(r)\to\infty$ and $B(r)\to0$, and a different one has $f(r)\to 0$ and $B(r)\to\infty$. In between these behaviours, by tuning $r_0$ one can find a solution that becomes asymptotically flat for a specific value $r_0=r_\star$. Such a solution is shown in Figure \ref{fig:wormholes}.
Comparing the $r\to\infty$ asymptotic behaviour obtained by numerical calculation to the asymptotically-flat case of the linearised theory solution \eqref{eqa:linearisedgensol} with $\beta=0$, one finds such an integral-step solution corresponding to $M\sim -C_{2,0}<0$ and $C_{2-}<0$.

To see why such a solution may be described as a wormhole, consider $f(r_0)=0$ and $B(r_0)=b_0>0$ and expand in $(r-r_0)$:
\be
f(r)=(r-r_0) f'(r_0) + \cdots\,,\qquad\notag
B(r)=b_0 + B'(r_0) (r-r_0) + \cdots\,.\label{wormholeasymp}
\ee
As one can see from the left-hand calculated plot in Fig.\ \ref{fig:wormholes}, both $B'(r_0)$ and $f'(r_0)$ are positive.
Now make a coordinate change in the radial coordinate according to 
\be
r-r_0=\ft14 \rho^2\,,\label{rhocoordchange}
\ee
after which one has the asymptotic form of the metric
\be
ds^2 = -(b_0+ \ft14 B'(r_0) \rho^2)dt^2 + \fft{d\rho^2}{f'(r_0)} +
(r_0^2 + \ft12 r_0 \rho^2) d\Omega^2 \ +\ldots\,.\label{wormmetric}
\ee
Since this solution is an even function of $\rho$, it is naturally $\Z_2$ symmetric in $\rho$ and can be extended to the full range $-\infty<\rho<+\infty$. Geodesics in the $\rho>0$ patch match smoothly onto geodesics for $\rho<0$ and so continue on naturally through to negative $\rho$ without hitting a singularity.

The interpretation as wormholes of the more general non-$\Z_2$-symmetric solutions with $f(r_0)=0$, $B(r_0)=\text{const.}$ arising from asymptotic expansions with half-integral steps in $(r-r_0)$, \ie in the $(w,t)=(1, 0)_{\sqrt{r-r_0}}$ family, is less clear. As we have seen, such solutions have four parameters in the expansion about $r_0$, which is the generic number for spherically symmetric solutions in the $\gamma R -\alpha C^2$ theory. This could allow tuning of one parameter combination so as to ensure asymptotic flatness at spatial infinity, even for an arbitrary value of $r_0$.  The expansion in half-integral powers of $(r-r_0)$, however, leads to odd powers of $\rho$ after making the coordinate change \eqref{rhocoordchange}. This destroys the $\Z_2$ symmetry of the $r_0$-tuned solution and invites the question whether one will then have $B\to 0$, and consequently a horizon, at some value $r_{\text{hor}}$ of the radius. Accordingly, the interpretation of such general $\Z_2$-asymmetric solutions as wormholes is not so clear as for the $\Z_2$ symmetric solutions.

\subsection{Schwarzschild and non-Schwarzschild black holes}\label{ssec:blackholesnum}

Turning now to asymptotically-flat solutions including a horizon, we know from the trace equation no-hair theorem of Section \ref{sec:tracenohair} that all such solutions must have vanishing Ricci scalar, $R=0$. Accordingly, the analysis of such solutions can be restricted to the $\gamma R-\alpha C^2$ theory, since the field equations of the general $\gamma$, $\alpha$, $\beta$ theory \eqref{HDGaction}  reduce to those of the $\beta=0$ theory when $R=0$. Furthermore, the results of Sections \ref{sec:horizonexp} and  \ref{sec:Schlinnohair} show that the Schwarzschild solution is in general {\em isolated} in the sense that the linearised no-hair theorem of Section \ref{sec:Schlinnohair} does not permit, for $0<\zeta = \alpha (\gamma r_0^2)^{-1}<\zmax$, where $r_0$ is the horizon radius, any solution infinitesimally deviating from Schwarzschild in the single non-Schwarzschild parameter allowed by parametric analysis near the horizon, as explained in Section \ref{sec:horizonexp}. As presented in Ref.\ \cite{Lu:2015cqa}, however, the qualified nature of the linearised no-hair theorem led to a suspicion that there might in fact be other asymptotically-flat solutions with horizons that are in general distinctly separated from Schwarzschild in the value of their non-Schwarzschild parameter. In the notation of Ref.\ \cite{Lu:2015cqa}, the non-Schwarzschild parameter may be taken to be $\delta$, defined by 
\bea
f(r) &=& \frac1{A(r)} = f_1\, (r-r_0) + {\cal O}(r-r_0)^2\notag\\
f_1 &=& \fft{1+\delta}{r_0}\,.\label{nonSchwparam}
\eea
The limiting value $\zmax$ of the linearised no-hair theorem suggests the existence of a branch point in the value of the horizon radius $r_0=r_0^{\rm min}$ at which infinitesimal values of $\delta$ {\em can} give rise to non-Schwarzschild asymptotically-flat solutions with horizons. As found in Ref.\ \cite{Lu:2015cqa}, for $r_0> r_0^{\rm min}$ such non-Schwarzschild solutions do exist, but they lie outside the  linear validity range of the $\delta$ parameter expansion. So, except at the $r_0^{\rm min}$ branch point, the Schwarzschild solution must be isolated. 

\begin{figure}[h]
\centering
\includegraphics[scale=.75]{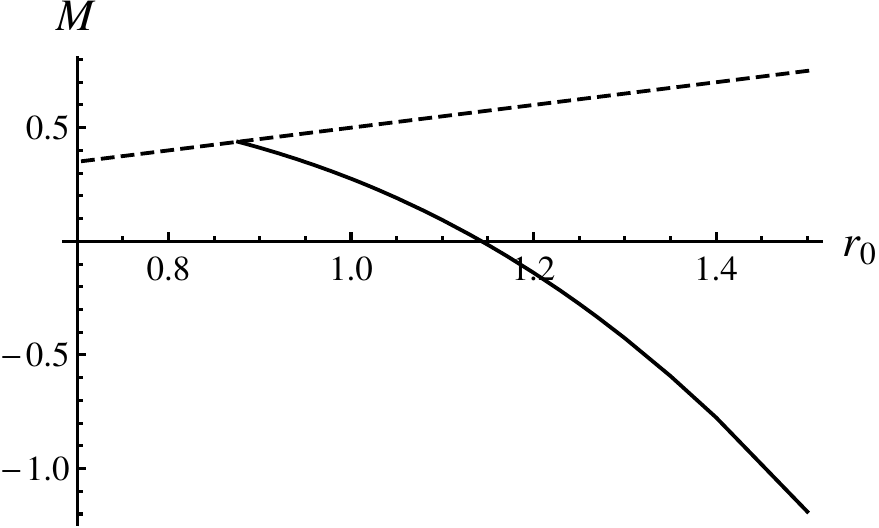}
\caption{Phase structure of the Schwarzschild (dashed line) and
non-Schwarzschild (solid line) black holes in a theory with $\alpha=\ft12$, sketched for a theory with $G=(16\pi\gamma)^{-1}=1$. The Schwarzschild mass is given by $GM=\frac12 r_0$.}
\label{fig:bhphases}
\end{figure}

Numerical calculations reveal the properties of the various black-hole phases. The phase structure \cite{Lu:2015cqa} in terms of black-hole mass $GM$ (as above, given \cite{ADMmass} by $\ft12$ the coefficient of $1/r$ in $g_{tt}$ as $r\to\infty$) is shown in Figure \ref{fig:bhphases}.

\begin{figure}[h]
\centering
\includegraphics[scale=.75]{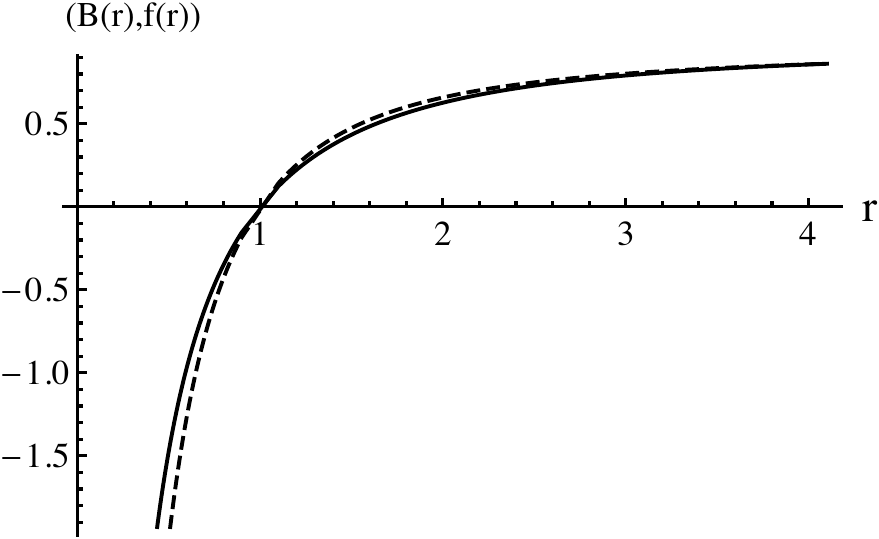}
\caption{Non-Schwarzschild black hole for $GM\sim 16 \pi \; 0.276$ with a horizon at $r=1$. The dashed line denotes $B(r)$ and the solid line denotes $f(r)=1/A(r)$.}
\label{fig:bhinsideout}
\end{figure}

Joining the non-Schwarzschild $r>r_0$ solution outside the horizon to the $r<r_0$ interior solution, one obtains the result shown in Figure \ref{fig:bhinsideout}.
The deviation from a Schwarzschild black hole is shown by the fact that $f(r)=1/A(r)\ne B(r)$; however, the approach as $r\to 0$ shows this solution can still fit naturally into the $(1,-1)$ indicial family. Another piece of information obtained from the calculation producing Figure \ref{fig:bhinsideout} is the sign of the $C_{2\,-}$ coefficient of the falling ${e^{-m_2r}\over r}$ spin-two Yukawa term. Both for positive and negative $GM=-\frac12 C_{2,0}$, the sign of $C_{2\,-}$ appears to be negative for such solutions. For $M>0$, this sign is opposite to that which would have been expected from the linearised $\gamma R - \alpha C^2$ theory coupled to a positive-energy shell delta-function source, as shown in \eqref{Vpoint} or \eqref{Vshell}. 

The numerical results presented in this section are clearly only an initial foray into the perhaps rich phase structure of the solution space of theories derived from the action \eqref{HDGaction}, and this subject clearly requires more careful numerical analysis.

\section{Conclusion}\label{sec:conclusion}

In this paper, we have carried out an analysis of the static spherically-symmetric solutions of the field equations derived from the action \eqref{HDGaction}. This extends older results \cite{Stelle:1977ry} by a full asymptotic analysis of the indicial $(s,t)=(0,0),\  (1,-1)$ and $(2,2)$ solution families  near the origin, together with a careful count of the parameters occurring in each family. The difficult question is then what happens in the intermediate $0\ll r\ll\infty$ region before one reaches spatial infinity, near which the assumption of asymptotic flatness once again allows for a closed-form study of solutions {\it via} the linearised version of the field equations derived from \eqref{HDGaction}. 

In Section \ref{sec:tracenohair}, we reviewed the no-hair theorem for the trace of the higher-derivative theory's field equations, agreeing with this part of the results of \cite{Nelson:2010ig}. This implies that asymptotically flat solutions containing a horizon must have $R=0$ throughout the whole extra-horizon spacetime. Consequently, one knows that asymptotically-flat solutions displaying a spin-zero Yukawa term (\ie with a non-zero coefficient $C_{0-}$) cannot have a horizon, since $C_{0-}\ne0 \implies R\ne0$. Study of solutions with vanishing Ricci scalar $R$, and hence without such a spin-zero Yukawa term, reduces to the somewhat simpler case of the $\gamma R-\alpha C^2$ theory with $\beta=0$. In this restricted context, the field equations can be reduced to a system of two coupled equations with at most second derivatives in $\partial\over \partial r$. Although there is no general no-hair theorem for the trace-free components of the field equations, one still does have a linearised no-hair theorem for the traceless field equations as derived in Section \ref{sec:Schlinnohair}. This implies that the Schwarzschild solution is generally isolated within the family of $(1,-1)$ solutions, which is controlled by a single non-Schwarzschild parameter, provided the horizon radius $r_0$ is larger than a certain bound $\sqrt{\alpha/(\gamma\zeta_{\text{max}})}$. As one moves away infinitesimally from the Schwarzschild solution within the $(s,t)=(1,-1)$ family, the only thing that can generally happen for solutions with a horizon is that asymptotic flatness is lost. 

What happens outside the domain of validity of the linearised no-hair analysis is another matter, however, and in Ref.\ \cite{Lu:2015cqa} asymptotically-flat non-Schwarzschild black-hole solutions were indeed found. Except near the $r_0\gtrsim \sqrt{\alpha/(\gamma\zeta_{\text{max}})}$ horizon radius bound for the linearised no-hair theorem, such non-Schwarzschild black holes can only exist owing to nonlinear dependence on the $(1,-1)$ family non-Schwarzschild parameter. Numerical evidence points to the $r_0\sim \sqrt{\alpha/(\gamma\zeta_{\text{max}})}$ horizon radius coinciding with a branch point in the black-hole solution space, at which the non-Schwarzschild black holes first occur (\cf Figure \ref{fig:bhphases}). This is clearly consistent with a breakdown of the linearised no-hair theorem for that radius, because just above such a branch point the non-Schwarzschild black holes should indeed be obtainable by linearised perturbation away from the Schwarzschild solution.

The general structure of the solution space for the higher-derivative \eqref{HDGaction} theory is still not completely clear, but we have found evidence of a rather rich phase structure for the static spherically symmetric and asymptotically flat solutions of this theory. We have seen from analysis of shell continuity and jump conditions that only the $(s,t)=(2,2)$ solution family (without horizons) of Section \ref{ssec:22solsnum} can couple to a standard distributional matter shell, but this need not imply that the $(2,2)$ family is the only equilibrium endpoint of gravitational collapse. In addition to the $(2,2)$ family, one has the Schwarzschild/non-Schwarzschild black-hole solutions of Section \ref{ssec:blackholesnum} plus the `wormhole' solutions of Section \ref{ssec:wormholesnum}. When the latter are $\Z_2$ symmetric, they lead cleanly through to a second sheet of spacetime, as in Eq.\  \eqref{wormmetric}. The asymptotically flat $(2,2)$ solutions, both Schwarzschild and non-Schwarzschild black holes, and also the $\Z_2$ wormhole solution all appear as separatrices between more generic singular solutions found numerically. The nature and interpretation of the latter remain to be better understood.

\subsection{Stability issues for black holes}\label{sec:stabilityissues}

Going beyond the static spherically-symmetric ansatz \eqref{eq:metric} is outside the scope of the present paper. But one can contemplate what could happen dynamically once time dependence is allowed. A full stability analysis of the various phases of the static solution space would be desired, but in the meantime one can extract some partial stability information from various quasinormal mode studies of the stability of the Schwarzschild solution itself, considered as a solution of the higher-derivative \eqref{HDGaction} theory. This has been studied, \egns, in Ref.\  \cite{Whitt:1985ki}. It was found there that the Schwarzschild solution is stable in the $\gamma R +\beta R^2$ theory with $\alpha=0$. This is not surprising, because that theory is equivalent \cite{Stelle:1977ry,Whitt:1984pd} to ordinary general relativity coupled to a positive-energy massive scalar field. 

In Ref.\  \cite{Whitt:1985ki} it was also suggested that the Schwarzschild solution could become unstable, for nontachyonic values of $(m_2)^2={\gamma\over2\alpha}$, for sufficiently small values of
\be
\mu_{\scriptscriptstyle W}={M m_2\over M_{\rm Pl}^2}\,,\label{mustability}
\ee
where $M_{\rm Pl}$ is the Planck mass. Ref.\  \cite{Whitt:1985ki} then went on to claim, nonetheless, that detailed analysis of the quasinormal modes of the theory \eqref{HDGaction} showed no such instability. This conclusion has, however, been challenged more recently in Ref.\ \cite{Myung:2013doa}, where it is claimed that Ref.\ \cite{Whitt:1985ki} erred in considering only a static S-wave potential instability. Instead, the analysis of Ref.\ \cite{Myung:2013doa} does find Schwarzschild S-wave instabilities for $\mu_{\scriptscriptstyle W}\lesssim 1$ by treating the Ricci tensor $R_{\mu\nu}$ as an effective massive field. This instability is compared to Schwarzschild instabilities found in massive theories of gravity \cite{massivegravSch}. 

Instability of the Schwarzschild solution for small black holes (\ie small $\mu_{\scriptscriptstyle W}$) raises the question whether a stable sector of the static solution space exists, and whether one or another of the non-Schwarzschild solutions we have discussed could then represent a stable final phase. Clearly, the relation between $\mu_{\scriptscriptstyle W}$ and the branch point in the black-hole solution space could be an important issue in this regard. The analysis of time-dependent gravitational collapse can, however, be quite involved, as indeed it is already for the apparently simpler case of Brans-Dicke theory \cite{ScheelSchapiroTeukolsky}. 

\section*{Acknowledgements}

We would like to thank Carl Bender, Alfio Bonnano, Stanley Deser and Toby Wiseman for helpful discussions during the course of the work. For hospitality during the course of the work, H.L., C.N.P. and K.S.S. thank the KITPC, Beijing;  H.L. and K.S.S. thank 
the Mitchell Institute, Texas A\&M; C.N.P. and K.S.S. thank the Cambridge-Mitchell Collaboration and the Mitchell Foundation for hospitality at the Great Brampton House workshop on Cosmology and Strings; and K.S.S. thanks the Perimeter Institute and the Institut des Hautes \'Etudes Scientifiques (for a visit supported in part by the National Science Foundation under Grant No. 1002477). The work of C.N.P.\ was supported in part by DOE grant  DE-FG02-13ER42020; 
the work of H.L.\ was supported in part by NSFC grants 11175269, 11475024 and 11235003; 
the work of K.S.S.\ was supported in part by the STFC under Consolidated Grant ST/J0003533/1 and the work of A.P.\ was supported by an STFC studentship. 
%\bigskip
\newpage

\addtocontents{toc}{\protect\newpage}

\section*{Appendices}
\begin{appendices}

\section{Reduced nonlinear field equations}\label{app:fullequations} 
The reduced field equations of maximal third order \eqref{eq:General} as derived in Section \ref{sec:differentialOrder} for spherically-symmetric solutions written in Schwarzschild coordinates are given in detail as follows, where $A^{(3)}$ and $B^{(3)}$ are third derivatives. For equation \eqref{eq:GeneralA} we have
\begin{align*}
24 r^4 A^3 B^4 H_{rr} =\ \,& 
8 r^3 A^2 B^2 B^{(3)} \left(r (\alpha -3 \beta ) B'-2 (\alpha +6 \beta ) B\right)
\\&
-4 r^2 A B^2 A'' \left(r^2 (\alpha -3 \beta ) B'^2-4 r (\alpha +6 \beta ) B B'+4 (\alpha -12 \beta ) B^2\right)
\\&
-4 r^4 (\alpha -3 \beta ) A^2 B^2 B''^2
\\&
-4 r^2 A B B'' \bigg(
	2 r B A' \left(
		r (\alpha -3 \beta ) B'-2 (\alpha +6 \beta ) B\right)
\\&
\ \ 	+A \left(
		3 r^2 (\alpha -3 \beta ) B'^2-12 r (\alpha +3 \beta ) B B'+8 (\alpha +6 \beta ) B^2\right)
	\bigg)
\\&
+7 r^2 B^2 A'^2 \left(
	r^2 (\alpha -3 \beta ) B'^2-4 r (\alpha +6 \beta ) B B'+4 (\alpha -12 \beta ) B^2\right)
\\&
+2 r^2 A B A' B' \left(
	3 r^2 (\alpha -3 \beta ) B'^2-4 r (2 \alpha +3 \beta ) B B'+4 (\alpha +24 \beta ) B^2\right)
\\&
+24 A^3 B^3 \left(
	\gamma  r^3 B'+B \left(
		\gamma  r^2-12 \beta \right)
	\right)
\\&
+A^2 \bigg(
	7 r^4 (\alpha -3 \beta ) B'^4-4 r^3 (5 \alpha +12 \beta ) B B'^3
\\&
\ \ 	-4 r^2 (\alpha -48 \beta ) B^2 B'^2+32 r (\alpha +6 \beta ) B^3 B'-16 (\alpha -21 \beta ) B^4 \bigg)
\\&
+8 A^4 B^4 \left(
	2 \alpha -6 \beta -3 \gamma  r^2\right)\,,\numberthis\label{eq:GeneralAdet}
\end{align*}
while for equation \eqref{eq:GeneralBB4B3}, using the definitions of $X(r)$ and $Y(r)$ as given in \eqref{X&Y}, we have
\begin{align*}
2 r^4 &A^5 B^2 \left(
	\alpha  r B'-3 \beta  r B'-2 \alpha  B-12 \beta  B\right)^2\Big(H_{tt} -X(r)H_{rr} -Y(r) \partial_r H_{rr}\Big) = 
\\& 
\ 72 \alpha  \beta  r^3 A^2 A^{(3)} B^4 \left(
	r (\alpha -3 \beta ) B'-2 (\alpha +6 \beta ) B\right)
\\&
+36 \alpha  \beta  r^2 A B^3 A''\bigg(
	13 r B A' \left(
		2 (\alpha +6 \beta ) B-r (\alpha -3 \beta ) B'\right)
\\&\ \ 	-2 A (
		-r^2 (\alpha -3 \beta ) B'^2+r (\alpha +6 \beta ) B B'+2 (\alpha +6 \beta ) B^2)
	\bigg)
\\&
+12 \beta  r^4 (\alpha -3 \beta ) A^3 B^2 B''^2 \left(
	(\alpha +6 \beta ) B-r (\alpha -3 \beta ) B'\right)
\\&
+4 r^3 A^2 B B'' \bigg(
	3 \beta  B A' \left(
		r^2 (\alpha -3 \beta )^2 B'^2+r \left(
			\alpha ^2-15 \alpha  \beta +36 \beta ^2\right)
		 B B'-6 \alpha  (\alpha +6 \beta ) B^2\right)
\\&
\ \ 	-3 \beta  A B' \left(
		-r^2 (\alpha -3 \beta )^2 B'^2-6 \alpha  r (\alpha -3 \beta ) B B'+2 \left(
			7 \alpha ^2+48 \alpha  \beta +36 \beta ^2\right)
		B^2\right)
\\&
\ \ 	+\gamma  (-r) (\alpha -3 \beta ) A^2 B^2 \left(
		2 (\alpha +6 \beta ) B-r (\alpha -3 \beta ) B'\right)
	\bigg)
\\&
+504 \alpha  \beta  r^3 B^4 A'^3 \left(
	r (\alpha -3 \beta ) B'-2 (\alpha +6 \beta ) B\right)
\\&
-3 \beta  r^2 A B^2 A'^2 \bigg(
	r^3 (\alpha -3 \beta )^2 B'^3+3 r^2 \left(
		17 \alpha ^2-57 \alpha  \beta +18 \beta ^2\right)B B'^2
\\&
\ \ 	-60 \alpha  r (\alpha +6 \beta ) B^2 B'-4 \left(
		23 \alpha ^2+150 \alpha  \beta +72 \beta ^2\right)
	B^3 \bigg)
\\&
-6 \beta  r A^2 B A' \bigg(
	r^4 (\alpha -3 \beta )^2 B'^4+r^3 \left(
		11 \alpha ^2-39 \alpha  \beta +18 \beta ^2\right)
	B B'^3-4 r^2 \left(
		8 \alpha ^2+51 \alpha  \beta +18 \beta ^2\right)B^2 B'^2
\\&
\ \ 	+4 r \left(
		11 \alpha ^2-12 \alpha  \beta +18 \beta ^2\right)
	B^3 B'-16 \left(
		4 \alpha ^2+21 \alpha  \beta -18 \beta ^2\right)
	B^4 \bigg)
\\&
+A^3 \bigg(
	-4 r (\alpha -3 \beta ) B^4 B' \left(
		12 \beta  (5 \alpha +3 \beta )+r (\alpha -3 \beta ) A' \left(
			\gamma  r^2-12 \beta \right)
		\right)
\\&
\ \ 	-2 r^2 B^3 B'^2 \left(
		6 \beta  \left(
			\alpha ^2+66 \alpha  \beta +36 \beta ^2\right)
		+\gamma  r^3 (\alpha -3 \beta )^2 A'\right)
\\&
\ \ 	-8 (\alpha +6 \beta ) B^5 \left(
		-6 \beta  (5 \alpha +3 \beta )-r A' \left(
			2 \alpha  \left(
				\gamma  r^2-6 \beta \right)
			+3 \beta  \left(
				12 \beta +\gamma  r^2\right)
			\right)
		\right)
\\&
\ \ 	-3 \beta  r^5 (\alpha -3 \beta )^2 B'^5+3 \beta  r^4 \left(
		-19 \alpha ^2+51 \alpha  \beta +18 \beta ^2\right)
	B B'^4+12 \beta  r^3 \left(
		13 \alpha ^2+84 \alpha  \beta +36 \beta ^2\right)
	B^2 B'^3 \bigg)
\\&
-8 A^5 B^4 \bigg(
	r (\alpha -3 \beta ) B' \left(
		\alpha  \left(
			\gamma  r^2-6 \beta \right)
		+6 \beta  \left(
			3 \beta +\gamma  r^2\right)
		\right)
\\&
\ \ 	+(\alpha +6 \beta ) B \left(
		\alpha  \left(
			6 \beta -2 \gamma  r^2\right)
		-3 \beta  \left(
			6 \beta +\gamma  r^2\right)
		\right)
	\bigg)
\\&
-2 A^4 B^2 \bigg(
	\gamma  r^5 (\alpha -3 \beta )^2 B'^3-6 r^2 (\alpha -3 \beta ) B B'^2 \left(
		\alpha  \left(
			\gamma  r^2-4 \beta \right)
		+3 \beta  \left(
			4 \beta +\gamma  r^2\right)
		\right)
\\&
	+4 r (\alpha -3 \beta ) B^2 B' \left(
		\alpha  \left(
			\gamma  r^2-24 \beta \right)
		+6 \beta  \left(
			\gamma  r^2-6 \beta \right)
		\right)
	+4 \left(
		2 \alpha ^2+15 \alpha  \beta +18 \beta ^2\right)
	B^3 \left(
		12 \beta +\gamma  r^2\right)
	\bigg)\,.\numberthis\label{eq:GeneralBB4B3det}
\end{align*}

From these two coupled third-order differential equations, one anticipates that the solution will depend in general on a total of six integration constants. For a pair of linear differential equations, this can be demonstrated straightforwardly by reducing the system to a single sixth-order differential equation for just one function, \eg $A(r)$, by repeatedly differentiating and substituting between equations so as to eliminate $B(r)$ and its derivatives. In the present highly nonlinear equation system (\ref{eq:GeneralAdet},\,\ref{eq:GeneralBB4B3det}), this is not feasible to do explicitly because this would involve the inversion of polynomials of $\mbox{order} >4$.  However, the idea can be outlined as a sequence of operations on Equations (\ref{eq:GeneralAdet},\,\ref{eq:GeneralBB4B3det}) as follows:
\begin{IEEEeqnarray}{rCl}
(\ref{eq:GeneralAdet}): 0 & = &\, f_1(r,A,B,A',B',A'',B'',B''')\notag\\
(\ref{eq:GeneralBB4B3det}): 0 & = &\, g_1(r,A,B,A',B',A'',B'',A''')\notag\\
\partial_r(\ref{eq:GeneralBB4B3det}): 0 & = &\, \partial_r g_1(r,A,B,A',B',A'',B'',A''')\notag\\
 &=& g_2(r,A,B,A',B',A'',B'',A''',B''',A^{(4)})\notag\\
\therefore B''' & = &\, g_2^{-1}(r,A,B,A',B',A'',B'',A''',A^{(4)})\notag\\
\text{sub into}\  (\ref{eq:GeneralAdet}): 0 & = &\, f_2(r,A,B,A',B',A'',B'',A''',A^{(4)})\notag\\
\therefore B'' & = &\, f_2^{-1}(r,A,B,A',B',A'',A''',A^{(4)})\notag\\
\text{sub into}\  (\ref{eq:GeneralBB4B3det}): 0 & = &\, g_3(r,A,B,A',B',A'',A''',A^{(4)})\notag\\
\therefore B' & = &\, g_3^{-1}(r,A,B,A',A'',A''',A^{(4)})\notag\\
\text{sub }f_2^{-1}\text{ and }g_3^{-1}\text{ into}\ (\ref{eq:GeneralBB4B3det}): 0 & = &\, g_4(r,A,B,A',A'',A''',A^{(4)})\notag\\
\therefore B & = &\, g_4^{-1}(r,A,A',A'',A''',A^{(4)})\notag\\
\text{sub into } (\ref{eq:GeneralBB4B3det}): 0 & = &\, g_5(r,A,A',A'',A''',A^{(4)},A^{(5)},A^{(6)})\,.
\end{IEEEeqnarray}

\section{Coupling of a shell source to the higher-derivative theory}\label{app:shellcoupling}
\subsection{Coupling an $(0,0)$ vacuum inside the shell to a $(2,2)$ solution outside}
One can carry out a successful coupling of a thin-shell delta-function stress-tensor source to the full nonlinear higher-derivative theory in a fashion similar to the couplings in the linearised theory as discussed in Section \ref{sec:weakFieldExpansion}. As we saw in Section \ref{sec:nonlinshellsources}, it's only with an exterior $(2,2)$ family solution combined with an interior $(0,0)$ family vacuum solution that the count of available interior plus exterior solution parameters is sufficient to satisfy the nine continuity, step, asymptotic flatness and asymptotic Minkowskian requirements. In this appendix, we discuss in more detail how these coupling requirements can be met.

Coupling a thin shell source to an external $(2,2)$ family solution faces a number of challenges. Principal among these is the lack of a closed-form expression for the $(2,2)$ family of solutions, so one is limited to carrying out the coupling using series solutions as given in Section \ref{sec:solutionsNearTheOrigin}. There is also an awkwardness arising from the choice of the Schwarzschild form \eqref{eq:metric} for the metric. Such difficulties were already noted in the classic treatment of delta-function couplings in general relativity given in Ref.\ \cite{Arnowitt:1960zzb}. Such difficulties might be alleviated by working in other than Schwarzschild coordinates, but we have not explored this possibility in detail. The difficulty with the $\text{in}\leftrightarrow\text{out}$ matching in the series-expanded theory is a leading-term matching problem in $A$: the inner solution has $A(0)=1$ and so we would naively expect $A(\ell_-) \approx 1 + O(\ell^2)$ but the outer solution has $A(\ell_+) \approx c_2 \ell^2 + O(\ell^3)$, while $A$ is continuous so $A(\ell_-)$ and $A(\ell_+)$ need to be equal for arbitrary $\ell$.  The $\ell \to 0$ limit is particularly interesting because this will give us the field of a point source in the higher-derivative theory. In order to make this work for arbitrary small $\ell$ we need to elevate the free parameters of the solution to functions of $\ell$.

Let us first illustrate the method with general relativity, where such dependence on the shell size is well-known \cite{Arnowitt:1960zzb}.

\subsection{Shell coupling in general relativity}
To set the stage for the more involved coupling problem in the higher-derivative theory, we first review the analogous coupling problem for distributional sources in general relativity. Some classic references for this are \cite{Arnowitt:1960zzb} and \cite{Geroch:1987qn}.

\subsubsection{Shell-source coupling using the closed-form Schwarzschild solution}
Consider general relativity with a thin-shell source as in (\ref{eq:theSource}--\ref{TthetaTtrel}), with $T_{tt} = \frac{M}{4 \pi \ell^2} \delta(r-\ell)$. It is convenient to define the length scale $\LM := 2 G M = M(8 \pi \gamma)^{-1}$.
Coupling the source to the equations of motion, one can show that $B$ is continuous while $A$ has a step:
\begin{align}
B_{\text{out}}(\ell_+) = &\, \ B_{\text{in}}(\ell_-)\ ,
&
A_{\text{out}}(\ell_+) - A_{\text{in}}(\ell_-) = &\, \ 
\frac{\LM A_{\text{in}}(\ell_-)^2}{\ell \, B(\ell) - \LM A_{\text{in}}(\ell_-)}\,.
\end{align}
In terms of these parameters the field of a spherical shell is
\bse
\begin{align}
A_{\text{in}} = &\, 1\,, \\
B_{\text{in}} = &\, b \,,\\
A_{\text{out}} = &\, \frac{1}{1-\frac{\LM}{b\, r}} \label{eq:GRanalyticSolA1m1}\,,\\
B_{\text{out}} = &\, \frac{b}{1-\frac{\LM}{b\, \ell}}\left(1-\frac{\LM}{b \, r}\right)\,, \label{eq:GRanalyticSolB1m1}
\end{align}\ese
where the exterior solution has the form of the Schwarzschild solution
\bse
\begin{align}
A_{\text{out}} = &\, \frac{1}{1-\frac{r_s}{r}}\,, \\
B_{\text{out}} = &\, k^2 \left(1-\frac{r_s}{r}\right)\,,
\end{align}\ese
provided the interior free parameter $b$ scales and the Schwarzschild radius $r_s$ is related to the source length $\LM$ via the $\ell$-dependent expressions
\bse
\begin{align}
b = &\, k^2 \left(1-\frac{r_s}{\ell}\right)\,, \label{eq:GRanalyticSolb}\\
\LM = &\, k^2 \left(1-\frac{r_s}{\ell}\right) r_s \,.\label{eq:GRanalyticSollM}
\end{align}\ese
Before proceeding with the core of our discussion, the signs here need a comment. 
For the case $0 < \ell < r_s$ a horizon exists and $M$ and $b$ are negative. 
It is familiar fact that we have $-+++$ signature outside the Schwarzschild horizon ($r > r_s$), and signature $+-++$ inside the Schwarzschild horizon ($\ell < r < r_s$). 
At the shell source ($r=\ell$), the function $B$ is continuous and non-vanishing so it is therefore positive inside and out. 
The function $A$, however, has a step at $r=\ell$. 
Outside the source (for $\ell < r < r_s$), $A$ is of course negative, but inside the shell source (for $r < \ell$) the equations of motion in the Schwarzschild-coordinate metric ansatz (where the angular part of the $ds^2$ metric is just $r^2 d\Omega^2$) require $A=1$. 
In consequence, the signature inside the shell is $++++$. 
The fact that the source mass $M$ has opposite sign to that of the Schwarzschild radius $r_s$ is not unexpected -- the source is static in a region where $t$ is spacelike, \ie the source is of tachyonic sign.
In the higher-derivative case this peculiarity will not arise, because we shall find that there is no horizon in the source-coupled solution, and the metric components $A$ and $B$ are continuous across the shell source. 
We thus anticipate a $-+++$ signature for all $r$ in the higher-derivative theory.

A key point in the above analysis is the fact that in general relativity the interior free parameter $b$ has to blow up as $\ell^{-1}$ as the shell is shrunk down to a point. We shall find that the higher-derivative case also requires poles at $\ell=0$  in the free parameters of the interior solution. 

\subsubsection{The Schwarzschild solution from a series-solution point of view}
To set the scene for analysis of coupling in the higher-derivative theory, where series solutions will be all that we have available, let us now repeat the above coupling discussion for the Einstein-theory Schwarzschild solution using only a series solution. The interior series vacuum solution is of $(0,0)$ structure:
\bse
\begin{eqnarray}
A_{(0,0)} & = &\, 1 + \dots\\
B_{(0,0)} & = &\, b + \dots
\end{eqnarray}\ese
and the solution exterior to the source is of $(1,-1)$ structure:
\bse
\begin{eqnarray}
A_{(1,-1)} & = &\, x r- x^2r^2+ x^3r^3- x^4r^4 + O(r^5) \\
B_{(1,-1)} & = &\, \frac{y}{r}+x y + \dots\,.
\end{eqnarray}\ese
To implement the matching conditions, we need to elevate the free parameters ($x,y,b$) to functions of $\ell$: ($x(\ell),y(\ell),b(\ell)$).

In order for the solution exterior to the source to be of the $(1,-1)$ family, we let the free parameters be expressed as Taylor series:
\bse
\begin{align}
x(\ell) = &\, x(0)+\ell x'(0)+\frac{1}{2} \ell^2 x''(0)+\dots \label{eq:GRSeriesSolx}\\
y(\ell) = &\, y(0)+\ell y'(0)+\frac{1}{2} \ell^2 y''(0)+\dots \label{eq:GRSeriesSoly}\,.
\end{align}\ese
For the interior free parameter $b$, however, we need to use a Laurent series -- \ie we allow $\ell^{-n}$ poles:
\begin{align}
%b = &\, \ell^a \left(b(0)+b'(0) \ell +\frac{1}{2} b''(0) \ell^2 +\dots\right) 
b = &\, \ell^a \left(b_0+b_1 \ell+b_2 \ell^2+b_3 \ell^3+\dots\right)
\label{eq:GRSeriesSolb}\,.
\end{align}
Then, in order for $y(\ell)$ to remain finite in the $\ell \to 0$ limit, we find that we need to scale $L_M$ (\ie we scale the mass $M$) as well:
\begin{align}
%\LM = &\, \ell^d \left(\LM(0)+\LM'(0) \ell_+\frac{1}{2} \LM''(0) \ell^2+\dots\right) 
\LM = &\, \ell^d \left(L_0+\ell L_1+\ell^2 L_2+\ell^3 L_3+\dots\right)
\label{eq:GRSeriesSollm}.
\end{align}
We find that the suitable poles have $a=-1 , \; d=-1$.

The solution is:
%\bse
%\begin{IEEEeqnarray}{rl}
%y = \ & 
%-\LM(\ell)
%\bigg(
%\ell x(0)
%+\ell^2 \left(x'(0)-x(0)^2\right)
%+O\left(\ell^3\right)
%\bigg)\notag
%\\
% =\  & 
%-\LM(0) x(0)
%-\ell \left(x(0) \LM'(0)+\LM(0) \left(x'(0)-x(0)^2\right)\right)
%+O\left(\ell^2\right)
%\\
%b =\  & -\frac{\LM(\ell) x(\ell)}{\ell} + O\left(\ell^3\right)\notag
%\\
% =\  &
%-\frac{\LM(0) x(0)}{\ell}
%-\left(x(0) \LM'(0)+\LM(0) x'(0)\right)\notag\\
%&\ \ \ -\frac{\ell }{2}\left(x(0) \LM''(0)+2 \LM'(0) x'(0)+\LM(0) x''(0)\right)
%+O\left(\ell^2\right)\,,
%\end{IEEEeqnarray}\ese
\bse
\begin{align}
y(\ell) = \ & 
-\LM(\ell)\left(\ell x(\ell)-\ell^2 x(\ell)^2 + O\left(\ell^3\right) \right)
\notag\\
 =\  & 
-L_0 x(0) + \ell \left(L_0 \left(x(0)^2-x'(0)\right)-L_1 x(0)\right)
 +O\left(\ell^2\right)
\\
b(\ell) =\  & -\LM(\ell) x(\ell) + O\left(\ell^3\right)\notag
\\
 =\  &
-\frac{L_0 x(0)}{\ell}
+\left(-L_0 x'(0)-L_1 x(0)\right)
\notag\\&
+\frac{1}{2} \ell \left(-L_0 x''(0)-2 L_1 x'(0)-2 L_2 x(0)\right)
+O\left(\ell^2\right)\,,
\end{align}\ese
which matches the analytic solution (where we hold $r_s$ fixed, something we don't know how to do in a series solution) for small $\ell$ upon renaming $x(0)=-\frac{1}{r_s}$ and 
$\LM(0)=-k^2 r_s^2$, and which also can be solved by matching for all $\ell$, producing
$x(\ell)= -\frac{1}{r_s}$ and $\LM(\ell)=k^2 \left(1-\frac{r_s}{\ell}\right) r_s$.

\subsection{Shell-source coupling in the higher-derivative theory}
Now we consider the coupling of a thin-shell stress-tensor source of the form (\ref{eq:theSource}--\ref{TthetaTtrel}) to the higher-derivative theory \eqref{HDGaction}. Inside the shell source, we require a $(0,0)$ vacuum solution, and outside the shell we consider a $(2,2)$ solution with the following notation:
\bse\begin{align}\label{eq:aroundOrigin22}
A = &\,\  r^2 w_2+\frac{r^3 v_3 w_2}{v_2}-\frac{r^4 \left(w_2 \left(2 v_2 \left(v_2 w_2-4 v_4\right)+v_3^2\right)\right)}{6 v_2^2}+r^5 w_5
+O(r^6)\,, \\
B = &\,\  r^2 v_2+r^3 v_3+r^4 v_4+r^5 v_5
+O(r^6)\,.
\end{align}\ese

The form (for at least $n \leq 14$) of the $(0,0)$ solution is
\bse
\begin{align}
%---------------------------------------------------
A = &\,
1
+a_2 r^2
+\sum_{n,p,q,m}
X_{n,p,m,q} \;\;
r^n
\left( \frac{\gamma}{\beta} \right)^{\frac{n}{2}-p}
a_2^{m}
\bTwoTilde^{p-m}
\left( \frac{\beta}{\alpha} \right)^{q}
%---------------------------------------------------
\\
\frac{B}{b_0} = &\,
1
+\bTwoTilde r^2
+\sum_{n,p,q,m}
Y_{n,p,m,q} \;\;
r^n
\left( \frac{\gamma}{\beta} \right)^{\frac{n}{2}-p}
a_2^{m}
\bTwoTilde^{p-m}
\left( \frac{\beta}{\alpha} \right)^{q}
%---------------------------------------------------
\end{align}\ese
where the $X_{n,p,m,q}$ and $Y_{n,p,m,q}$ are rational numbers and the $n,p,q,m$ sums are taken over $n=4,6,8,\ldots$; $1 \leq p \leq \frac{n}{2}$; $0 \leq q \leq \frac{n}{2}-1$ and $0 \leq m \leq p$.

Similarly to the method used for general relativity, (\ref{eq:GRSeriesSolx}), (\ref{eq:GRSeriesSoly}), (\ref{eq:GRSeriesSolb}) and (\ref{eq:GRSeriesSollm}), 
we elevate the free parameters to functions of $\ell$ in the following scheme
\begin{equation}
\begin{aligned}
b_0 = &\, \ell^2 H_0(\ell)
\,,&&&
w_2 = &\, w_2(\ell)
\,,\\
a_2 = &\, \ell^{-2} G_2(\ell)
\,,&&&
v_2 = &\, v_2(\ell)
\,,\\
\bTwoTilde = &\, \ell^{-2} F_2(\ell)
\,,&&&
v_3 = &\, v_3(\ell)
\,,\\
\,&&&&
v_4 = &\, v_4(\ell)
\,,\\
\,&&&&
w_5 = &\, w_5(\ell)
\,,\\
M   = &\, \ell^d \mu (\ell)
\,,&&&
v_5 = &\, v_5(\ell)\,,
\end{aligned}
\end{equation}
where we have factored out the poles so the remaining functions 
are Taylor series in $\ell$.

To carry out matching across the source shell, we need to have $A_{\text{in}}(\ell_-) = A_{\text{out}}(\ell_+)$, where
\begin{equation}
A_{\text{in}}(\ell_-) = 
1
+ G_2(\ell)  
+ \sum_{k,n,q,m} \ell^k \left( \frac{\gamma}{\beta} \right)^{\; \half k} X_{n,\frac{n-k}{2},m,q} \; 
G_2(\ell)^{m}
F_2(\ell)^{\frac{n-k}{2}-m}
\left( \frac{\beta}{\alpha} \right)^q\,,
\end{equation}
where the sum is taken over $k=0,2,4,6,\dots$; $k+2 \leq n=4,6,8,\dots$; $0 \leq q \leq \frac{n}{2}-1$ and $0 \leq m \leq \frac{n-k}{2}$.
Gathering powers of $\ell$, we have $A_{\text{in}}(\ell_-) \sim \ell^0 + \sim \ell^1 + \sim \ell^2 + \dots$ and $A_{\text{out}}(\ell_+) \sim \ell^2 + \dots$, so the leading-term matching problem in $A$ is now displayed in the vanishing of the interior $\ell^0$ term, where one has for $A_0$ (the $\ell^0$ term in $A_{\text{in}}$)
\begin{equation}\label{ell_0term}
A_0 = 
1+ G_2(0)+ \sum_{n,q,m}
\; X_{n,\frac{n}{2},m,q} \; 
G_2(0)^m F_2(0)^{\frac{n}{2}-m} \; 
\left(\frac{\beta}{\alpha}\right)^{q}\,,
\end{equation}
where the sum is taken for $n=4,6,8,\dots$; $0 \leq q \leq \frac{n}{2}-1$ and $0\leq m\leq\frac{n}{2}$.
The higher terms in \eqref{ell_0term} need to decrease in amplitude with $n$ so that the sum converges, and its limiting value as $n \to \infty$ must vanish so as to match the structure of $A_{\text{out}}$. We have not carried out an exhaustive analysis of the convergence properties of the resulting series, but we may consider the structure in the simplifying limit $\beta << \alpha$. In this limit, the $q \geq 1$ terms are suppressed and we need only consider $X_{n,\frac{n}{2},m,0}$. In this limit, the numbers $X_{n,\frac{n}{2},\frac{n}{2},0}$ are equal to 1 for all  $n$ while the other $X_{n,\frac{n}{2},m\leq \frac{n}{2}-1,0}$ grow at most linearly with $n$ (for at least $n \leq 14$). If we rename $t=\frac12 n$, $X_{n,\frac{n}{2},m,0}=X_{t,m}$ and $G_2(0)=\zeta F_2(0)$ and take the sum only out to $T$ terms, we have
\be
A_0 = 
\sum_{\substack{t=1,2,3,4,...T\\ 0 \leq m \leq t}}
\; X_{t,m} \; 
\zeta^m  F_2(0)^t\,.
\ee
Given the at most linear growth in $t$ of $X_{t,m}$, one gets an estimate of the sum $A_0$ by letting $X_{t,m} = a+bt$, for which an estimate sum $\tilde A_0$ can be carried out:
\be
\tilde A_0 = \sum_{t=1,2,3,4,...T}
\; (a + b t) \; 
\frac{1-\zeta^{t+1}}{1-\zeta}  F_2(0)^t\,,
\ee
for which the ratio of successive terms at large $t$ is 
\be
\frac{a+b+bt}{a+bt} \,\;
\frac{F_2(0)^{t+1}}{F_2(0)^t} \,\;
 \frac{1-\zeta^{t+2}}{1-\zeta^{t+1}} \,
\sim
F_2(0)\,.
\ee
Accordingly, convergence is obtained in the  $\beta << \alpha$ limit. Convergence is expected when the $\beta << \alpha$ limit is relaxed as well.

Once the $\ell^0$ terms are matched inside and out, the matching of $\ell^{N \geq 1}$ terms and the matching of $A'(\ell), B(\ell),B'(\ell),B''(\ell), A''(\ell)$ should follow suit with less difficulty. The final result will describe the interior and exterior metrics in terms of $\alpha, \beta, \gamma, \LM(\ell)$ and three other free parameters $p_1 (\ell), p_2(\ell), p_3(\ell)$.

\section{On higher-derivative no-hair theorems}\label{app:Nelson}
In this appendix, we present a recalculation of the argument of Ref.\ \cite{Nelson:2010ig} including a cosmological constant. Unfortunately, this corrects the calculation of \cite{Nelson:2010ig} in a way that voids its conclusion about the no-hair consequences of the non-trace part of the higher-derivative gravity field equation. 

The discussion of \cite{Nelson:2010ig} can be generalised to the case with a cosmological constant. Take as Lagrangian
\begin{equation}
I = \int d^4x\sqrt{-g}\left(\gamma (R-2\Lambda) -
\alpha C_{\mu\nu\rho\sigma}C^{\mu\nu\rho\sigma} + \beta R^{2}\right)\,.
\end{equation}
The equation of motion (\ref{generalEOM}) gains a term 
\begin{equation}
H_{\mu\nu} \to H_{\mu\nu} + \gamma \Lambda g_{\mu\nu}\,,
\end{equation}
making the trace
\begin{eqnarray}
H_{\mu}^{\;\mu} & = &\, 6\beta \left(\Box R - m_0^2 (R-4\Lambda)\right)
\\
& = &\, 6\beta \left(\Box S - m_0^2 S\right)\,,
\end{eqnarray}
where we have defined 
\begin{equation}
S = R - 4 \Lambda \,.
\end{equation}
For the trace no-hair theorem, the discussion proceeds from this point on just as in the case without a cosmological constant as reviewed in Section \ref{sec:tracenohair}, obtaining finally
\begin{equation}
0=\int_{\cal S} \sqrt{h}\; {\rm d}^3x\left[
 D^a \left(\lambda\, S\; D_a S\right) -\lambda\,
\left( D^a S \right)\left( D_a S\right) -  \lambda \; m_0^{\;2}  \, S^2 
\right]\,.
\end{equation}
The outer boundary contribution vanishes if $D_a S = D_a R$ falls off appropriately fast at infinity. Consequently, we deduce that 
\be
S=0 \quad\Leftrightarrow\quad R=4\Lambda\label{Svanishes}
\ee 
if the inner boundary term vanishes, which is ensured if the inner boundary is at a horizon.

The argument for the traceless part of the higher-derivative no-hair theorem runs into trouble, however.

Define the shifted quantities
\be
S_{\mu\nu} = R_{\mu\nu}-g_{\mu\nu}\Lambda\,, \qquad
S = g^{\mu\nu}S_{\mu\nu} = R - 4 \Lambda\,.
\ee
Letting $\threeR$ denote the Ricci scalar of the spatial part of the metric $h_{ab}$, we also define the shifted quantity
\begin{equation}
\threeS = \threeR -2\Lambda
\end{equation}
and finally we define $m_2(\Lambda)$ such that $m_2(0)=m_2$
\begin{equation}
m_2(\Lambda)^2 := \frac{\gamma+\frac{8}{3}\Lambda(3\beta-\alpha)}{2 \alpha}\,,
\end{equation}
thus obtaining the equations of motion
\begin{equation}\label{eq:NelsonThm3EOMS0}
0 = \left. \frac{H_{\mu\nu}}{-2 \alpha} \right|_{S=0} = 
\Box S_{\mu\nu}-m_2(\Lambda)^2 S_{\mu\nu}
+2 S_{\mu}^{\rho}S_{\nu\rho} - 2 \nabla_{\rho}\nabla_{\mu}S_{\nu}^{\rho}
-\frac{1}{2}g_{\mu\nu}S^{\rho\sigma}S_{\rho\sigma}\,,
\end{equation}
which smoothly analytically continue from $\Lambda=0$ to $\Lambda$ finite.
Note that the contracted Bianchi identity in this case is
\begin{align}
\nabla^{\mu} R_{\mu\nu} = &\, \half \nabla_{\nu} R\,,
\\
\therefore\  \nabla^\mu S_{\mu\nu} = &\, 0 \quad \text{for}\ S=0\,,
\\
\nabla^\mu S_{\mu i } = 0 = &\, D^j S_{ij} + \frac{1}{\lambda} S_{ij}D^j \lambda + \frac{\threeS}{2\lambda} D_i \lambda\,.
\end{align}
At this point, we can contract (\ref{eq:NelsonThm3EOMS0}) with $\lambda\;S^{\mu\nu}$, dimensionally reduce on the $t$ coordinate, and integrate over three-dimensional space while using \eqref{Svanishes} to obtain
\bse\label{eq:NelsonThm3}
\begin{align}
0 = & \int_{\cal S} \sqrt{h}\; {\rm d}^3x \left[ \left. \lambda\;S^{\mu\nu}\frac{H_{\mu\nu}}{-2 \alpha} \right|_{S=0} \right] 
\\
 =
 & \int_{\cal S} \sqrt{h}\; {\rm d}^3x \Bigg[
D_i \left(
\frac{\lambda}{4}\threeS D^i \threeS + \lambda\;S^{kl} D^i S_{kl} - 2 \lambda\;S_{kl}D^{k}S^{li} - \lambda\;\threeS D_jS^{ji}
\right)
\\&
-\frac{\lambda}{4}\;D^i \threeS \; D_i \threeS + 2\lambda\;D^i \threeS D^jS_{ji}
-\lambda\;D^iS^{jk} \left[D_iS_{jk}-2D_jS_{ki}\right]\label{opsigns}
\\&
-\lambda\;\frac{\threeS^2}{4}\left(m_2(\Lambda)^2 + \threeS \right)
-\lambda\;S^{ij}S_{ij} \left(m_2(\Lambda)^2-2\mathcal{S}\right)
\Bigg]\,.
\end{align}\ese

Unfortunately, owing to the fact that the squared terms in the middle line \eqref{opsigns} of the integrand are of non-uniform sign, it is \emph{not} possible in the case of higher-derivative gravity to say that the assumption of a horizon and suitable boundary conditions at infinity imply that $S_{\mu\nu}$ vanishes, even for large $m_2(\Lambda)$. This continues smoothly to $\Lambda=0$, where (\ref{eq:NelsonThm3}) with a horizon and asymptotic flatness unfortunately does not imply that $R_{\mu\nu}$ vanishes,  even for large $m_2$.

\end{appendices}

\end{document}